\documentclass[10pt]{article}
\usepackage[square,authoryear]{natbib}
\usepackage{marsden_article}
\usepackage{epstopdf}
\usepackage{setspace}
\usepackage{bbm}
\usepackage{amscd}

\usepackage{caption2}

\DeclareGraphicsExtensions{.eps,.ps,.pdf}

\begin{document}

% THEOREM Environments ---------------------------------------------------
 \newtheorem{thm}{Theorem}[section]
 \newtheorem{cor}[thm]{Corollary}
 \newtheorem{lem}[thm]{Lemma}
 \newtheorem{prop}[thm]{Proposition}
 \newtheorem{defn}[thm]{Definition}
 \newtheorem{rem}[thm]{Remark}
 \numberwithin{equation}{section}
% MATH -------------------------------------------------------------------

\title{\Large{\textbf{A Geometric Theory of Thermal Stresses}}\footnote{To appear in the \emph{Journal of Mathematical Physics}.}}

\author{Arkadas Ozakin\thanks{Georgia Tech Research Institute, Atlanta, GA 30332.}
\and Arash Yavari\thanks{School of Civil and Environmental
Engineering,
  Georgia Institute of Technology, Atlanta, GA 30332. E-mail: arash.yavari@ce.gatech.edu.} }

%\date{September 29, 2005}
\maketitle

\begin{abstract}
In this paper we formulate a geometric theory of thermal stresses.
Given a temperature distribution, we associate a Riemannian
material manifold to the body, with a metric that explicitly
depends on the temperature distribution. A change of temperature
corresponds to a change of the material metric. In this sense, a
temperature change is a concrete example of the so-called
referential evolutions. We also make a concrete connection between
our geometric point of view and the multiplicative decomposition
of deformation gradient into thermal and elastic parts. We study
the stress-free temperature distributions of the
finite-deformation theory using curvature tensor of the material
manifold. We find the zero-stress temperature distributions in
nonlinear elasticity. Given an equilibrium configuration, we show
that a change of the material manifold, i.e. a change of the material
metric will change the equilibrium configuration. In the case of a
temperature change, this means that given an equilibrium
configuration for a given temperature distribution, a change of
temperature will change the equilibrium configuration. We obtain
the explicit form of the governing partial differential equations
for this equilibrium change. We also show that geometric
linearization of the present nonlinear theory leads to governing
equations that are identical to those of the classical linear
theory of thermal stresses.
\end{abstract}

%\begin{description}
%\item[Keywords:] Continuum Mechanics, Elasticity, Internal energy, Stress tensor, Differential forms.
%\end{description}

\tableofcontents

%\newpage

\section{Introduction}

% make the difference btw elasticity and plasticity more explicit
% stress-free: am i using this incorrectly? should i mention
% explicitylethat no ext stresses are...?

Classical elasticity theory quantifies the amount of stretch in a
body by using a specific configuration as the reference
configuration. The displacements as measured from the reference
configuration and the strains associated with them are then used
to get the stresses via constitutive relations. This viewpoint
works nicely when there is a relaxed, stress-free configuration
that can be used as the reference configuration. However, this is
not always the case. A body may have various sources of residual
stresses, e.g. defects such as dislocations and disclinations, in
which case there may not exist any stress-free configuration. One
can observe the existence of these residual stresses by cutting
pieces from the body when there are no external forces, and see
that the pieces relax upon being cut. One can deal with these
residual stresses in the classical theory \citep{BoleyWeiner1997}
but we will use a different approach in this paper; that of
geometric elasticity, and the notion of a space that describes the
intrinsic properties of a body with a residual stress
distribution.

Temperature enters free energy density as a state variable. In
classical linear theory of thermal stresses
\citep{BoleyWeiner1997}, it is assumed that there exists a
reference temperature $T_0$ at which the body is stress free. Free
energy is then expanded about $T_0$ and only linear and quadratic
terms are kept. The governing equations of this theory consist of
those of linearized elasticity and heat conduction with some
coupling terms. Given an equilibrium configuration of the body at
temperature $T$, a change in temperature will change the
equilibrium configuration due to the coupling terms. Similar ideas
are used in the nonlinear theory by looking at thermal stresses as
a coupled nonlinear elasticity/heat conduction problem.

In this paper, we study thermal stresses geometrically by
considering a material manifold that explicitly depends on
temperature. Material manifold is endowed with a Riemanian metric
$\mathbf{G}$. We assume that given a reference temperature distribution $T_0$,
when the body is unloaded in a Riemannian manifold
$(\mathcal{B},\mathbf{G}_0)$ it is stress free. Change of
temperature changes the metric. For similar ideas in the case of
dislocations see \citep{BilbyBulloughSmith1955, Kondo1955a,
Kondo1955b, Kondo1963, Kondo1964, Kroner1992, Kroner1996}. We
should emphasize that here we assume that temperature distribution
is given. A geometric formulation of the coupled elasticity/heat
conduction will be discussed in a future communication.

As a motivation for our viewpoint, consider a piece from a thin,
elastic spherical shell being forced to lie on a plane, e.g., by
being squeezed between two flat surfaces. This constraint will
induce stresses on the shell. Let us for the moment imagine that
we are observing this shell from the two-dimensional viewpoint of
the plane, ignoring the third dimension. When we cut pieces from
the shell (all still forced to lie on the plane), we will observe
that the former relax by a certain amount, demonstrating the
``residual stress'' on the body. The two-dimensional, planar
viewpoint dictates that there is no stress-free configuration for
this piece of material. However, if this same shell is ``forced''
to live on the surface of a sphere of appropriate radius, there
will be a stress-free configuration, which may then be used as the
reference configuration for measuring the amount of stretch, etc.
The surface of the sphere is intrinsically two-dimensional, in the
sense that it can be described without any reference to a third
dimension, by using only two coordinates, and an intrinsic measure
of distance, e.g. by using the spherical coordinates:
%-----------------------------
\begin{equation}
    ds^2 = R^2 (d\theta^2 + \sin^2\theta~d\phi^2).
\end{equation}
%-----------------------------
This suggests a generalization to three dimensions: given a body
with residual stresses with no apparent stress-free configuration,
could it be possible to find some abstract three-dimensional space
in which it would be stress-free? As stated, this is a very
general question, and the answer depends on the source of the
residual stresses for the case at hand and the notion of ``space''
one is considering.

In this paper, we answer this question affirmatively for a
specific source for residual stresses: that induced by the thermal
expansion due to a non-uniform temperature distribution in a solid
material. Our analysis is constructive; given a temperature
distribution on a previously stress-free material, we construct a
Riemannian manifold (to be defined shortly) in which the material
under consideration with the given thermal profile would have a
stress-free state. This construction is not of purely academic
interest, since given the appropriate constitutive relations, it
allows us to calculate the stresses in an elastic body in a given
thermal setting, and is not restricted to a linear response. The
appropriate way to quantify deformations for a body with residual
stresses is by considering a map (``the configuration map'') from
the ``material manifold'' to the three-dimensional spatial space
that the body lives in. The constitutive relations are then
written in terms of this map.

In the remaining of the paper we study the effect of changes of
the material manifold on equilibrium configuration. We do this in
a general setting when the material manifold is Riemannian. We
also consider the case where the change in metric is small enough
so that a linear approximation would be enough. In the case of
small changes in material manifold we obtain the governing
equations of evolution of the equilibrium configuration as a
function of changes of the material manifold.

Let us pause here and explain the main ideas and conclusions.
Suppose we have a flat, two-dimensional elastic material on a
plane. If we heat this material in a non-uniform way, it will tend
to bend out from the plane, and take the shape of a curved
surface. If we then force this curved surface to live in the flat
plane by perhaps squeezing it between two flat, rigid surfaces, we
will induce some stresses on it. This suggests that for a given
non-uniform temperature distribution, there may be a curved,
stress-free shape that an ``originally flat" elastic material
wants to take. When we force this material to live on a flat
plane, we induce ``thermal stresses". Of course in real life, we
have three-dimensional elastic bodies, forever bound to live in
flat, three-dimensional Euclidean space.

Here is the main idea: given a material metric $\mathbf{G}$
describing the stress-free state of an elastic material at
constant temperature, and a spatial metric $\mathbf{g}$ (for
simplicity, both of these metrics can be taken as flat,
three-dimensional metrics), we claim that a non-uniform
temperature distribution on the material should be represented as
a change in the material metric $\mathbf{G}$. In particular, for a
material with isotropic thermal properties, we claim that the
required change in $\textbf{G}$ is just a pointwise rescaling
(i.e. multiplying the metric with a scalar function on the
material manifold), and the way the scaling factor depends on the
temperature is determined by the physical properties of the
material. Non-isotropic thermal expansion will also be considered.

We believe this is an example where a change in the material
manifold can be clearly understood conceptually. This may be very
helpful in understanding the role of the change in material
connection in defect mechanics, where one has to consider not only
a change in the metric, but also the ``torsion part" of the
connection. In particular, we have the hope that an analogy with
thermal expansion will help clarify the often encountered (and, in
our opinion, confusing) discussions of decomposing the deformation
gradient $\mathbf{F}$ into ``elastic" and ``plastic" parts. The
analogue of a plastic deformation in our case is a change in the
temperature, resulting in a change in the material connection.
Such a change in the material connection may induce stresses on a
previously stress-free configuration, and as a result, change the
equilibrium configuration. The resultant stresses in the new
equilibrium configuration should be explored with the ``usual"
geometric elasticity.

This paper is structured as follows. In \S2 we motivate the
connection between thermal stresses and changes in material
metric. We do this by looking at the example of a two-dimensional
disk and study the possibility of existence of a relaxed
configuration embedded in the three-dimensional Euclidean space
for an arbitrary radial temperature distribution. We also study
the stress-free temperature distributions in nonlinear elasticity.
In \S3 we present the main ideas of a geometric formulation of
thermal stresses. We study the effect of a change of material
manifold on the equilibrium configuration. In \S4 the geometric
linearization of the nonlinear theory is presented. Conclusions
are given in \S5. To make the paper self contained, we briefly
review the basic concepts of differentoial geometry,
parralelizable manifolds, and geometric theory of elasticity in
the appendix.

\section{The Material Metric and Non-Uniform Temperature Distributions}

\paragraph{Motivation.} Suppose we start with a stress-free isotropic material %% elastic, not
%% plastic. also: isotropic
with a uniform temperature distribution $T_1$ and free boundary
conditions, and increase its temperature to
$T_2$.  The material will expand, %%?
and the original distance $\delta L_1 $ between two neighboring
points $A$ and $B$ in the solid body will increase to $\delta
L_2$. The quantity $(\delta L_2 - \delta L_1) / \delta L_1$ turns out to be %%better
%%argument?
independent of the two points, i.e., the expansion is uniform.
Note that
%-----------------------------
\begin{equation}
    \frac{\delta L_2-\delta L_1}{\delta L_1}=\alpha (T_2-T_1),
\end{equation}
%-----------------------------
where $\alpha$ is the coefficient of thermal expansion. Let us now
assume that we use a Lagrangian coordinate system, $X^1, X^2,
X^3$,\footnote{The superscripts denote coordinate labels.}i.e.,
assume that the same material points have the same coordinates
before and after the expansion. Then, the distance between the two
points $A$ and $B$ is given approximately in terms of the metric
tensor $G_{ij}$ as follows:
%-----------------------------
\begin{equation}
    \delta l \approx \sqrt{G_{ij}(X^i_B - X^i_A)(X^j_B-X^j_A)},
\end{equation}
%-----------------------------
where the components $G_{ij}$ are evaluated at a point between
$X_A$ and $X_B$. This shows that $G_{ij}$ should somehow depend on
temperature. In other words, this suggests that, in this
Lagrangian setting, we should be using different metric tensors
for $T_1$ and $T_2$. Note that this relation between the
Lagrangian coordinates  and the material manifold works only
because the material is in a relaxed, stress-free state in both
temperatures. Otherwise, the distance that the material metric
$\mathbf{G}$ measures would not be the spatial distance.

\paragraph{The thermal expansion coefficient.} Let us now connect the
above description of the material metric in terms of a temperature
distribution to the thermal expansion coefficient used in the
classical theory. We imagine a material with a nonuniform
temperature distribution, for various values of the constant
temperature. The thermal expansion coefficient is defined by
looking at the equilibrium volume of the material at different
temperatures. Let us therefore look at the volume element of a
material at a given constant temperature. We assume that the
material metric is given by
%-----------------------------
\begin{equation}\label{material-metric}
  G_{IJ}(\mathbf{X},T) = H_{IJ}(\mathbf{X}) e^{2 \omega(T)},
\end{equation}
%-----------------------------
where $H_{IJ}$ is independent of temperature and
$T=T(\mathbf{X})$. The volume form associated with this metric is
given as
%-----------------------------
\begin{equation}
  dV(\mathbf{X},T) = \sqrt{\det{|H_{IJ}|}} ~e^{N \omega(T)} d^NX,
\end{equation}
%-----------------------------
where $d^NX$ is shorthand for $dX^1 \wedge dX^2 \cdots \wedge
dX^N$.\footnote{Note that $N=2$ or $3$.} Differentiating with
respect to $T$, we obtain
%-----------------------------
\begin{equation}
  \frac{d}{dT}dV(\mathbf{X},T) = \sqrt{\det{|H_{IJ}|}}~e^{N \omega(T)} d^NX ~\frac{Nd\omega(T)}{dT}= dV(\mathbf{X},T) \,N\,\frac{d\omega(T)}{dT}.
\end{equation}
%-----------------------------
Thus, we can read-off the thermal expansion coefficient in terms
of the temperature dependence of $\omega(T)$:\footnote{Note that
for thermally isotropic materials volumetric thermal strain is $N$
times the thermal strain.}
%-----------------------------
\begin{equation}\label{coefficient-thermal-expansion}
  \alpha(T) = \frac{d\omega(T)}{dT}\,.
\end{equation}
%-----------------------------

\paragraph{Remark:} Suppose $H_{IJ}=\delta_{IJ}$, i.e. the
initial material manifold is Euclidean. In this case $G_{IJ}$ is
conformally flat. It is known that any 2-dimensional Riemannian
manifold is conformally flat and the function $\omega$ is unique
\citep{Berger2003}. See the appendix for more discussions on this.

\subsection{Stress-Free Temperature Distributions}

Before we go into the general geometric theory of thermal
stresses, we will first demonstrate an important application of
our geometric approach, that of stress-free temperature
distributions.

\paragraph{The two-dimensional case.} As in the introduction, consider a
two-dimensional shell restricted to live on a flat planar surface
between two rigid planes. We will assume that initially the shell
is at constant temperature, and is stress-free, with no external
or body forces. We would like to find the temperature
distributions that will result in equilibrium configurations with
zero stress. Changing the temperature uniformly will result in
uniform expansion, and hence no stress. Are there other
temperature distributions with this property? The answer is yes as
is already known in the framework of linear thermoelasticity
\citep{BoleyWeiner1997}. Due to the nature of our geometric
approach, we will also be able to answer this question in a
nonlinear setting. To the best of our knowledge, this has not been
done in the literature.

The spatial distances between material points are measured by the
ambient space metric (the ``spatial metric''), which is Euclidean.
A given temperature distribution will result in a change in the
material metric, as described above. A configuration will be
stress-free if there is no ``stretch'' in the material, i.e., if
the material distance between two points is the same as the
spatial distance. This happens only if the two types of metric
tensors (spatial and material) agree on the distance measurements
between nearby material points, i.e. only if they are isometric. Since the spatial metric is
assumed to be Euclidean, this means that the material metric,
after the change due to a given thermal distribution, has to be
Euclidean.

It is worth emphasizing that one cannot simply set $G_{IJ} =
\delta_{IJ}$, the precise requirement is that the pull-back of the
spatial (Euclidean) metric by the deformation map $\varphi$ has to
be equal to the material metric. This issue is closely related to
the fact that a metric may be Euclidean ``in disguise'', i.e., one
can write the flat two-dimensional metric in different coordinate
systems, and it is not always easy to recognize that the metric is
flat by simply looking at its components in a given coordinate
system.

Riemann's original work solves this problem for any dimensionality
by defining the curvature tensor of the metric: a metric is flat,
i.e., it can be brought into the Euclidean form $\delta_{IJ}$ {\it
locally} by a coordinate transformation, if and only if its
curvature tensor is zero \citep{Berger2003}.\footnote{We have
collected the definitions of various curvature-related quantities
of interest in the appendix.}

It turns out that in two dimensions, a weaker requirement is
sufficient \citep{Berger2003}: a metric is flat if and only if its
scalar curvature (the Ricci scalar) is zero. Let us now apply this
condition to a two-dimensional metric that is obtained from a
non-uniform temperature distribution on an initially stress-free,
planar shell, i.e., $G_{IJ}=e^{2\Omega}\delta_{IJ}$, where
$\Omega(\mathbf{X}) = \omega(T(\mathbf{X}))$. The Ricci scalar for
a metric of this form is given by \citep{Wald1984}
%-----------------------------
\begin{equation}
  \textsf{R} = -2\,e^{-2\Omega} \nabla^2\Omega\,.
\end{equation}
%-----------------------------
Thus, $\textsf{R}=0$ requires $\nabla^2\Omega=0$, i.e., the
exponent in the scale factor has to be a harmonic function. If we
assume that $\omega(T)$ depends on temperature {\it linearly}, we
obtain $\nabla^2 T=0$. This is exactly the same condition
encountered in linear elasticity, see, e.g.
\cite{BoleyWeiner1997}. This means that for the case of constant
thermal expansion coefficient in two dimensions, harmonic
temperature distributions do not result in any stresses. However,
our result is more general: even if the thermal expansion is
non-linear, we obtain the condition $\nabla^2\Omega(X) = \nabla^2
\omega(T(X)) = 0$, where $\omega(T)$ gives the general, non-linear
dependence of (isotropic) thermal expansion to temperature.

It is worth emphasizing the distinction between local and global
flatness, and the implications for stress-free thermal
distributions. Even though the surface of a right circular
cylinder in three dimensions looks curved, it is locally, {\it
intrinsically} flat. For any given point on the cylinder, one can
find a finite-sized region containing the point, and a
single-valued coordinate system on this region, for which the
metric has the Euclidean form. Physically, this means that for any
given point, we can cut some finite-sized piece containing the
point, and can lay the piece on a flat plane, without stretching
it. The surface of a sphere in three dimensions, on the other
hand, is intrinsically curved; it is impossible to make any
finite-sized piece of the sphere, no matter how small, to lie on a
flat plane without stretching it.

Curvature conditions like $\textsf{R}=0$, or $\nabla^2 \Omega=0$
can only detect such local issues. That it is impossible to make a
full cylinder lie in a plane nicely (i.e., without tearing,
folding, or stretching it) is due to the global topology of the
cylinder, and local restrictions on curvature are not capable of
constraining the global properties sufficiently.

\begin{figure}[hbt]
\begin{center}
\includegraphics[scale=0.6,angle=0]{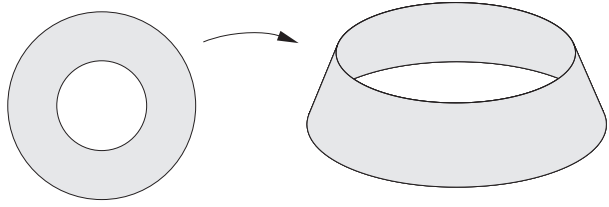}
\end{center}{\vskip -0.1 in}
\caption{\footnotesize Zero-stress deformation of an annulus to a
cone.} \label{Cone}
\end{figure}

In the context of thermal stresses, this subtlety is nicely
demonstrated by the following example. Let us specialize to the
case where $\Omega$ depends only on the radial coordinate $R$ of
an initially flat annular piece of a material, $R_0\le R \le R_1$.
The flatness condition gives
%-----------------------------
\begin{equation}
  \nabla^2 \Omega = \frac{1}{R}\frac{d}{dR}\left(R\,\frac{d\Omega(R)}{dR}\right) = 0\,.
\end{equation}
%-----------------------------
Solving this gives
%-----------------------------
\begin{equation}
  \label{radial-stress-free}
  e^{2\Omega} = \gamma R^{2\beta},
\end{equation}
%-----------------------------
where $\gamma>0$ and $\beta$ are constants.\footnote{Assuming
$T(R_0)
  = T_0$, and that the coefficient of thermal expansion is a constant
  $\alpha$, this solution corresponds to the temperature distribution
%-----------------------------
\begin{equation}\label{two-d-log-temp}
  T(R)=T_0+\frac{2\beta}{\alpha}\ln \left(\frac{R}{R_0}\right).
\end{equation}
%-----------------------------
} Thus, we are concerned with temperature distributions that
result in metrics of the form
%-----------------------------
\begin{equation}
  dS^2 = R^{2\beta}\left(dR^2 + R^2\,d\Theta^2\right),
\end{equation}
%-----------------------------
where we have set $\gamma=1$ by a rescaling of $R$. It may not be
immediately obvious that these metrics are flat, but a
transformation to a new radial coordinate $r$ by $R =
r^{1\over\beta +1}$ gives
%-----------------------------
\begin{equation}
  dS^2 = c^2 dr^2 + r^2 d\Theta^2,
\end{equation}
%-----------------------------
where $c=\frac{1}{\beta+1}$. A further transformation,
$\tilde{R}=|c| r$, $\tilde{\Theta} = \Theta/|c|$ gives
%-----------------------------
\begin{equation}
  dS^2 = d\tilde{R}^2 + \tilde{R}^2
  d\tilde{\Theta}^2\,,\label{cone-metric}
\end{equation}
%-----------------------------
which is the flat two-dimensional metric, except for an important
subtlety. In the original coordinates, a point $(R,\Theta)$ was
identified with the point $(R, \Theta + 2\pi)$. In terms of the
new coordinates, this means that a point $(\tilde{R},
\tilde{\Theta})$ needs to be identified with the point
$(\tilde{R}, \tilde{\Theta} + 2\pi/|c|)$. The geometric meaning of
this is clear: the metric (\ref{cone-metric}), with the proper
identifications, is describing an annular piece from a conical
surface, with deficit angle $\alpha = 2\pi(1-1/|c|)$, see
Fig.~\ref{Cone}.

%-----------------------------
%-----------------------------
%-----------------------------
%-----------------------------

\begin{figure}[hbt]
\begin{center}
\includegraphics[scale=0.6,angle=0]{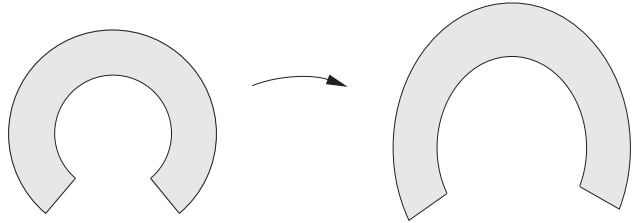}
\end{center}{\vskip -0.1 in}
\caption{\footnotesize Zero-stress deformation of a
simply-connected piece of an annulus.} \label{Annulus}
\end{figure}

Now, intuitively, one can guess that it will be impossible to make
such a conical surface lie on the plane without tearing,
stretching, or folding it. Thus, if we start with an annular shell
between two rigid planes, a temperature distribution of the form
(\ref{two-d-log-temp}) will indeed {\it result in stresses},
although the related material metric is intrinsically flat.
However, if the material consists only of a simply-connected piece
of the annulus (say, $R_1<R<R_2$,
$0<\Theta_1<\Theta<\Theta_2<2\pi$), the temperature distribution
(\ref{two-d-log-temp}) will just cause a stress-free expansion of
the material, between the two rigid planes. See Fig.
\ref{Annulus}.
%-----------------------------
%-----------------------------
%-----------------------------
%-----------------------------

This issue exists in the classical theory of elasticity, as well,
see p. 256 in \citep{BoleyWeiner1997} where a set of global
conditions have been given for stress-free thermal distributions
in two-dimensions for multiply-connected bodies. As an application
of these ideas, suppose we are given a temperature distribution
$T(\mathbf{X})$, and are seeking a temperature-dependent coefficient of
thermal expansion $\alpha(T)$ that would result in a stress-free
equilibrium state. In general, this amounts to solving the
equation
%-----------------------------
\begin{equation}\label{zero-temperature}
  \nabla^2\Omega=\frac{d\alpha}{dT}\frac{\partial T}{\partial X^A}\frac{\partial T}{\partial X^B}\delta^{AB}+\alpha\frac{\partial^2 T}{\partial X^A \partial X^B}\delta^{AB}
  =\frac{d\alpha}{dT}\left |\nabla T \right|^2+\alpha \nabla^2 T = 0.
\end{equation}
%-----------------------------
This equation may or may not admit a solution, depending on
$T(\mathbf{X})$, but for the simple case of a radial temperature
distribution $T(R)$, the solution (\ref{radial-stress-free})
dictates that $\alpha(T)$ is given by
%-----------------------------
\begin{equation}
  \alpha(T) = \frac{d\omega}{dT} = \frac{\frac{d\Omega}{dR}}{\frac{dT}{dR}} = \frac{\beta}{R\,T'(R)}\,.
\end{equation}
%-----------------------------
As above, however, this works only for simply-connected pieces
such as those described by  $\Theta_1<\Theta<\Theta_2$ and
$R_1<R<R_2$, with $\Theta_2 - \Theta_1 < 2\pi$. With this in mind,
$\alpha(T)$ for some specific radial temperature distributions are
as follows.
%-----------------------------
 \begin{eqnarray}
   T(R)=\frac{T_0R_1-T_1R_0}{R_1-R_0}+\left[\frac{T_1-T_0}{R_1-R_0}\right]R &:& \alpha(R)=\alpha_0\frac{R_0}{R}, \\
   T(R)=\frac{T_1R_1-T_0R_0}{R_1-R_0}+\left[\frac{R_0R_1(T_0-T_1)}{R_1-R_0}\right]\frac{1}{R} &:& \alpha(R)=\alpha_0\frac{R}{R_0}, \\
   \label{Constant_Alpha} T(R)=\frac{\ln\left(R_1^{T_0}/R_0^{T_1}\right)}{\ln\left(R_1/R_0\right)}+\left[\frac{T_1-T_0}{\ln\left(R_1/R_0\right)}\right] \ln R &:& \alpha(R)=\alpha_0.
 \end{eqnarray}
 %-----------------------------
% The solution for a general temperature distribution is
% %-----------------------------
% \begin{equation}
%   \alpha(R)=\alpha_0 e^{-\int_{R_0}^{R}\left[\frac{1}{r}+\frac{T''(r)}{T'(r)}\right]dr}.
% \end{equation}
% %-----------------------------

Let us next consider the case where the two-dimensional material
is allowed to bend into the third dimension, instead of being
squeezed between two rigid flat planes. Assume, once again, that
we start with a stress-free, planar piece of material at uniform
temperature, and we introduce a non-uniform temperature
distribution. Will the material be able to find a stress-free
state by bending into the third dimension? We will investigate
this problem for the special case of a radial temperature
distribution, and cylindrically symmetric configurations in three
dimensions.

According to our approach, the metric
%-----------------------------
\begin{equation}\label{radial-metric}
  dS^2 = e^{2\omega(T(R))}(dX^2 +
  dY^2) = e^{2\omega(T(R))} (dR^2 + R^2d\Theta^2),
\end{equation}
%-----------------------------
describes the natural, stress-free state of the shell, and for a
given configuration in the ambient 3-dimensional space, stresses
will be due to a different metric being induced on the shell by
the embedding in the ambient space. We will seek embeddings for
which the metric induced from the ambient space is the same as the
intrinsic metric, which, as opposed to the case considered above,
is not necessarily flat.

Let us begin by writing the induced metric for a given
configuration with cylindrical symmetry, which is best done in
cylindrical coordinates $(\rho, \phi, z)$. Instead of the most
general configuration, we will seek a solution of the form
%-----------------------------
\begin{eqnarray}
  \phi(R, \Theta) &=& \Theta,\\
  \rho(R,\Theta) &=& \rho(R),\\
  z(R,\Theta) &=& z(R)\,.
\end{eqnarray}
%-----------------------------
For such an embedding, the metric induced from the ambient space
is given as
%-----------------------------
\begin{equation}
  dS_{\textrm{induced}}^2 = dR^2\left[\left(\frac{dz}{dR}\right)^2 +
  \left(\frac{d\rho}{dR}\right)^2 + \rho(R)^2 \right].
\end{equation}
%-----------------------------
In order for this induced metric to be the same as the intrinsic,
material metric given by (\ref{radial-metric}), we need
%-----------------------------
\begin{eqnarray}
  \rho(R) &=& R\,e^{\Omega(R)},\\
  \left(\frac{dz}{dR}\right)^2 +   \left(\frac{d\rho}{dR}\right)^2 &=& e^{2\Omega(R)}\,,
\end{eqnarray}
%-----------------------------
where we define $\Omega(R) = \omega(T(R))$. Substituting $\rho(R)$
in the second equation, we obtain
%-----------------------------
\begin{equation}
  \left(\frac{dz}{dR}\right)^2 +
  e^{2\Omega(R)}\left[1+R\,\Omega'(R)\right]^2 = e^{2\Omega(R)}\,,
\end{equation}
%-----------------------------
which, in principle, lets us solve for $z(R)$. For a region where
$\rho(R)$ is invertible, we can also obtain the surface in three
dimensions as given by $z(\rho)$, by solving
%-----------------------------
\begin{equation}\label{z-rho}
  \left[1 + R\,\Omega'(R(\rho))\right]^2\left[1 + \left(\frac{dz}{d\rho}\right)^2\right] = 1\,.
\end{equation}
%-----------------------------
Note that for a uniform temperature distribution $T(R)=T_0$,
$\Omega'(R(\rho))=0$ and hence $z(\rho)=z_0$, which is what we
expect, i.e., in this case the relaxed configuration is planar.
Note also that (\ref{z-rho}) has a solution only if $-2/R<
\Omega'(R)<0$. In order to have some insight about the meaning of
this constraint, let us specialize to conical metrics described
above in (\ref{cone-metric}), $e^{2\Omega} = \gamma R^{2\beta}$.
For this case, the constraint $-2/R< \Omega'(R)<0$ translates to
$-2<\beta<0$, which gives, in terms of the deficit angle $\alpha$,
$0<\alpha<2\pi$. The condition $\alpha<2\pi$ is not surprising,
but there is nothing wrong with a cone with negative deficit angle
in terms of intrinsic geometry. The lower bound on $\Omega$ is
simply telling us that it is not possible to embed such a cone in
$\mathbb{R}^3$ in a cylindrically symmetric way, which makes
intuitive sense, considering the twisted shape of a saddle.

\paragraph{The three-dimensional case.}
Let us next consider the three-dimensional case. In three
dimensions, a vanishing Ricci scalar is not sufficient to
guarantee local flatness, however, a three-dimensional metric is
flat if and only if its Ricci tensor vanishes \citep{Berger2003}.
The Ricci tensor $\mathcal{R}_{IJ}$ of the metric $G_{IJ} =
e^{2\Omega}G^{(0)}_{IJ}$ is given in terms of the Ricci tensor
$\mathcal{R}^{(0)}_{IJ}$ of $G^{(0)}_{IJ}$ by the following
relation \citep{Wald1984}:
%-----------------------------
\begin{equation}
  \mathcal{R}_{IJ}=\mathcal{R}^{(0)}_{IJ} - (n-2)\nabla_I\nabla_J\Omega
  - G^{(0)}_{IJ}\left(G^{(0)}\right)^{KL}\nabla_K\nabla_L\Omega + (n-2)\nabla_I\Omega\nabla_J\Omega-
    (n-2)G^{(0)}_{IJ}\left(G^{(0)}\right)^{KL}\nabla_K\Omega\nabla_L\Omega\,,
\end{equation}
%-----------------------------
where $n$ is the dimensionality. Now, once again, assume that the
initial metric $G^{(0)}_{IJ} = \delta_{IJ}$,
$\mathcal{R}^{(0)}_{IJ}=0$, and $n=3$, and replace the covariant
derivatives with partial derivatives. This gives
%-----------------------------
\begin{equation}
  \mathcal{R}_{IJ} = -\partial_I\partial_J\Omega -
    \delta_{IJ}\delta^{KL}\partial_K\partial_L\Omega +
    \partial_I \Omega \partial_J\Omega - \delta_{IJ}\delta^{KL}\partial_K\Omega\partial_L\Omega=0.
\end{equation}
%-----------------------------

Similar to what is implicitly done in classical linear
thermoelasticity, let us assume that the reference temperature
$T_0$ is uniform, i.e. independent of position and that change of
temperature is ``small". This means that $\frac{\partial
T}{\partial X^I}$ is small. But note that
%-----------------------------
\begin{equation}
    \frac{\partial\Omega}{\partial X^I}=\alpha(T)\frac{\partial T}{\partial
    X^I}.
\end{equation}
%-----------------------------
Therefore, $\partial_I \Omega$ is small too, i.e. quadratic terms
in $\partial_I\Omega$ can be ignored. This gives us the condition
that all the second derivatives of $\Omega$ have to vanish. This
means that $\Omega$ is a linear function of the original Euclidean
coordinates.  If we further assume that $\omega(T)$ is a linear
function of temperature, we see that temperature itself has to be
a linear function of the original Euclidean coordinates.
Therefore, we recover the classical result in linearized thermal
elasticity that in three dimensions, the only stress-free
temperature distributions for an initially stress-free material
depend linearly on the coordinates \citep{BoleyWeiner1997}.

In the nonlinear case $\mathcal{R}_{IJ}=0$ is
equivalent to the following system of nonlinear partial
differential equations in terms of $\Omega$:
%-----------------------------
\begin{eqnarray}
  && \label{flatness-1}  \Omega_{,12}=\Omega_{,1}\Omega_{,2}, \\
  && \label{flatness-2}  \Omega_{,13}=\Omega_{,1}\Omega_{,3}, \\
  && \label{flatness-3}  \Omega_{,23}=\Omega_{,2}\Omega_{,3}, \\
  && \label{flatness-4}  \Omega_{,11}+\nabla^2\Omega+\Omega_{,2}^2+\Omega_{,3}^2=0, \\
  && \label{flatness-5}  \Omega_{,22}+\nabla^2\Omega+\Omega_{,1}^2+\Omega_{,3}^2=0, \\
  && \label{flatness-6}  \Omega_{,33}+\nabla^2\Omega+\Omega_{,1}^2+\Omega_{,2}^2=0.
\end{eqnarray}
%-----------------------------
Let us first look at the following nonlinear partial differential
equation for $w=w(x,y)$.
%-----------------------------
\begin{equation}
  \frac{\partial^2 w}{\partial x \partial y}=\frac{\partial w}{\partial x}\frac{\partial w}{\partial y}.
\end{equation}
%-----------------------------
Using the change of variable $u=e^{-w}$, one obtains
%-----------------------------
\begin{equation}
  \frac{\partial^2 u}{\partial x \partial y}=0.
\end{equation}
%-----------------------------
Thus, the most general solution is
%-----------------------------
\begin{equation}
  w(x,y)=-\ln[f(x)+g(y)],
\end{equation}
%-----------------------------
for some arbitrary functions $f$ and $g$. Using
(\ref{flatness-1})-(\ref{flatness-3}), one can show that
%-----------------------------
\begin{equation}\label{solution}
  \Omega(X^1,X^2,X^3)=-\ln[f(X^1)+g(X^2)+h(X^3)],
\end{equation}
%-----------------------------
for some arbitrary functions $f,g,$ and $h$. Now substituting
(\ref{solution}) into (\ref{flatness-3})-(\ref{flatness-6}), we
obtain
%-----------------------------
\begin{equation}
  f''(X^1)=g''(X^2)=h''(X^3).
\end{equation}
%-----------------------------
Therefore
%-----------------------------
\begin{eqnarray}
  && f(X^1)=c_0(X^1)^2+d_1X^1+d_2, \nonumber \\
  && g(X^2)=c_0(X^2)^2+d_3X^2+d_4, \\
  && h(X^3)=c_0(X^3)^2+d_5X^3+d_6, \nonumber
\end{eqnarray}
%-----------------------------
for some constants $c_0,d_1,...,d_6$.\footnote{Note that the case
$c_0=d_1=d_3=d_5=0$ corresponds to uniform temperature
distributions.} Plugging these back into
(\ref{flatness-4})-(\ref{flatness-6}), we get
%-----------------------------
\begin{equation}
  \Omega(X^1,X^2,X^3)=-\ln\left[c_0\sum_{i=1}^3(X^i-b^i)^2\right].
\end{equation}
%-----------------------------
Shifting the origin $X^i\to X^i+b^i$, this becomes
%-----------------------------
\begin{equation}
  \Omega(X^1,X^2,X^3)=-\ln\left(c_0R^2\right),
\end{equation}
%-----------------------------
where $R = \sqrt{(X^1)^2 + (X^2)^2 + (X^3)^2}$.

\begin{figure}[hbt]
\begin{center}
\includegraphics[scale=0.45,angle=0]{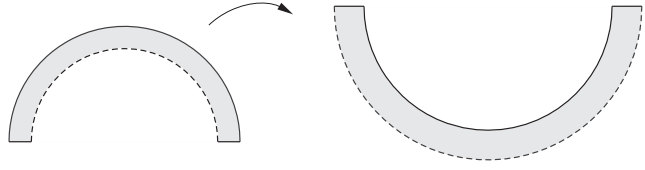}
\end{center}{\vskip -0.1 in}
\caption{\footnotesize A nonlinear stress-free thermal deformation
of a ball. Note that because of symmetry only deformation of a
great half circle of the projected sphere on a plane passing
thorugh the center of the ball is shown in this figure.}
\label{Zero-Stress}
\end{figure}

If $\alpha$ is constant, then this corresponds to the following
temperature distribution\footnote{Note that this is similar in
form to the 2D solution (\ref{Constant_Alpha}).}
%-----------------------------
\begin{equation}
  -\ln\left(c_0R^2\right)=\alpha(T-T_0)~~~\textrm{or}~~~T-T_0=c_0-\frac{2}{\alpha}\ln R.
\end{equation}
%-----------------------------
Note that $c_0$ represents a uniform change in temperature. In
order to understand what this solution represents physically, let
us write the metric in polar coordinates.
%-----------------------------
\begin{equation}
	dS^2 = e^{2\Omega} \left[dR^2 + R^2 (d\Theta^2 + \sin^2\Theta d \Phi^2)\right] 
       = \frac{1}{c^2R^4}\left[dR^2 + R^2 (d\Theta^2 + \sin^2\Theta d
       \Phi^2)\right].
\end{equation}
%-----------------------------
Now let us define
%-----------------------------
\begin{equation}  \label{inversion}
 \tilde{R} = \frac{1}{cR}.
\end{equation}
%-----------------------------
In terms of $\tilde{R}$, the metric becomes
%-----------------------------
\begin{equation}
  dS^2 = d\tilde{R}^2 + \tilde{R}^2(d\Theta^2 + \sin^2\Theta
  d\phi^2),
\end{equation}
%-----------------------------
which is precisely the flat Euclidean metric in three dimensions.
Thus, after the thermal expansion, the metric is still flat, but
the radial coordinate in which it is manifestly so is related to
the old radial coordinate by (\ref{inversion}) (up to a simple
shift of origin). This means that, particles at the two radii $R_1
< R_2$ move to the new radii $\tilde{R}_1 > \tilde{R}_2$, after
the thermal expansion, i.e., the material gets ``inverted''. This
may not be possible for a solid ball without tearing it apart, but
it is perfectly possible for a piece from such a ball, as
demonstrated in Fig.~\ref{Zero-Stress}.
%-----------------------------
%-----------------------------
%-----------------------------
%-----------------------------

\subsection{Connection with Multiplicative Decomposition of Deformation Gradient}

Geometric study of thermal stresses goes back to the works of
\cite{Stojanovic1964,Stojanovic1969}. These researchers extended
Kondo's \citep{Kondo1955a,Kondo1955b,Kondo1963,Kondo1964} and
Bilby's \citep{BilbyBulloughSmith1955,BilbyGardnerStroh1956} idea
of local elastic relaxation in the continuum theory of distributed
defects to the case of thermal stresses. See also \citet{Maugin2003} for a review of some relevant works. One should note that the idea of using differential geometry in anelasticity goes back to an earlier work by \citet{Eckart1948}.

Stojanovi\'c's idea is similar in spirit to the approach described
in this paper: a nonuniform temperature distribution in general
leads to residual stresses essentially because the body is
constrained to deform in Euclidean space. If one partitions the
body into small pieces, each piece will individually relax, but it
is impossible to realize a relaxed state for the full body by
combining these pieces in Euclidean space. Any attempt to
reconstruct the full body by sticking the particles together will
induce deformations on them, and will result in stresses. An
imaginary relaxed configuration for the full body is incompatible
with the geometry of Euclidean space.

The approach taken in this paper is to ask the question: which
space, as opposed to the Euclidean space, would be compatible with
a relaxed state of the body?
We claimed above that the answer to this question is, a Riemannian manifold
whose metric is related to the nonuniform temperature distribution
by (\ref{material-metric}). This metric describes the relaxed state
of the material, and the strains in a given configuration
should be measured with respect to the relaxed state, i.e., the new
material metric. In this setup, the constitutive relation (e.g., a
free energy function) is given
in terms of the material metric, the
(Euclidean) spatial metric, and the deformation gradient
$\mathbf{F}$. The constitutive relation allows one to
calculate the stresses induced
for a given configuration, at least in principle.

Stojanovi\'c, on the other hand, takes the following viewpoint.
Consider one of the imaginary relaxed pieces described above. The
process of relaxation after the piece is cut corresponds to a
linear deformation of this piece (linear, since the piece is
small)\footnote{This transformation is not necessarily uniquely
determined, see below for a discussion of this issue.}. Let us
call this deformation $\mathbf{F}_T$. If this piece is deformed in
some arbitrary way after the relaxation, one can calculate the
induced stresses by using the tangent map of this deformation mapping in the
constitutive relation.

Now, in order to calculate the stresses induced for a given
deformation of the {\it full body}, we focus our attention to one
such particular piece. The deformation gradient of the full body
at this piece $\mathbf{F}$ can be decomposed as $\mathbf{F} =
\mathbf{F}_e\mathbf{F}_T$, where, by definition, $\mathbf{F}_e =
\mathbf{F}\mathbf{F}_T^{-1}$. Thus, as far as this piece is
concerned, the deformation of the full body consists of a
relaxation, followed by a linear deformation given by
$\mathbf{F}_e$. The stresses induced on this piece, for an
arbitrary deformation of the full body, can be calculated by
substituting $\mathbf{F}_e$ in the constitutive relation.

Note that $\mathbf{F}_e$ and $\mathbf{F}_T$ are not necessarily
true deformation gradients in the sense that one cannot
necessarily find deformations $\varphi_{e}$ and $\varphi_{T}$ whose
tangent maps are given by $\mathbf{F}_e$ and $\mathbf{F}_T$,
respectively. This is due precisely to the incompatibility
mentioned above. However, as long as we have a prescription for
obtaining $\mathbf{F}_e$ and $\mathbf{F}_T$ directly for a given
deformation map $\varphi$ for the body and a temperature
distribution, we can calculate the stresses by the following
prescription.

For an isotropic material, Stojanovi\'c gives the following
formula for $\mathbf{F}_T$ in terms of the temperature.
%-----------------------------
\begin{equation}\label{ft-definition}
    (F_T)^A{}_B=\vartheta(T)\delta^A_B.
\end{equation}
%-----------------------------
This means that a small piece relaxes by a uniform expansion,
whose magnitude is determined by a function $\vartheta(T)$ that
characterizes the thermal expansion properties of the material
under consideration. Given this formula for $\mathbf{F}_T$, we can
calculate $\mathbf{F}_e = \mathbf{F}\mathbf{F}_T^{-1}$ for a given
deformation, and utilize a constitutive relation that gives the
stresses in terms of $\mathbf{F}_e$.

These two approaches seem very different philosophically, and at
first sight, $\mathbf{F}_T$, the ``incompatible intermediate
deformation gradient,'' perhaps seems a little mysterious from the
geometric standpoint. However, these approaches are related, as we
will demonstrate next. Our discussion can easily be generalized to
other sources of residual stresses; a local relaxation approach
and the Riemannian approach are equivalent for a large class of
settings.\footnote{However, there are cases that will require a
further generalization, namely, cases where the material manifold
has a connection with torsion and non-metricity.}

As mentioned above, Stojanovi\'c
\citep{Stojanovic1964,Stojanovic1969} gives
$(F_T)^A{}_B=\vartheta(T)\delta^A_B$ and relates the coefficient
of thermal expansion $\alpha$ to $\vartheta(T)$ by
%-----------------------------
\begin{equation}
    \alpha(T)=\vartheta(T)\frac{d \vartheta(T)}{dT}.
\end{equation}
%-----------------------------
This can agree with (\ref{coefficient-thermal-expansion}) only if
%-----------------------------
\begin{equation}\label{alpha-theta-relation}
    e^{\omega(T)}=\vartheta(T).
\end{equation}
%----------------------------
In order to show the mechanical equivalence of the two approaches
by using this identification, we need to show that for any given
constitutive relation for one of the approaches, one can find a
corresponding constitutive relation for the other approach that
predicts the same stresses for all possible deformations when
$\vartheta$ and $\omega$ are related through
(\ref{alpha-theta-relation}).

The constitutive relations of the two approaches are formulated in
terms of different quantities: $\mathbf{F}_e =
\mathbf{F}\mathbf{F}_T^{-1}$ on one side, and $\mathbf{G}(T)$ and
$\mathbf{F}$ on the other.  Let us start with our approach,
namely, assume that a constitutive relation is given in terms of
$\mathbf{G}(T)$ and $\mathbf{F}$. This
takes the form of a scalar free energy function that depends on
$\mathbf{G}(T)$, $\mathbf{F}$, as well as on the spatial metric tensor
$\mathbf{g}$, and possibly $\mathbf{X}$ and $T(\mathbf{X})$
explicitly:
%-----------------------------
\begin{equation}\label{constitutive-riemann}
  \Psi=\Psi(\mathbf{X},T,\mathbf{G}(\mathbf{X},T),\mathbf{F},\mathbf{g}).
\end{equation}
%-----------------------------

Now, $\mathbf{G}$, $\mathbf{F}$, and $\mathbf{g}$ are tensors,
written in terms of specific bases for the material space and the
ambient space. Commonly, bases associated to coordinate systems
are used. A change of basis changes the components of these
tensors, but $\Psi$, being a scalar, does not change. Let us
consider a change of basis from the original coordinate basis
$\mathbf{E}_A$ of the material space, satisfying
%-----------------------------
\begin{equation}
    \langle\!\langle \mathbf{E}_{A},\mathbf{E}_{B} \rangle\!\rangle_{\mathbf{G}}=G_{AB}\,,
\end{equation}
%-----------------------------
to an orthonormal basis $\hat{\mathbf{E}}_{\hat{A}}$ that
satisfies
%-----------------------------
\begin{equation}
    \langle\!\langle \hat{\mathbf{E}}_{\hat{A}},\hat{\mathbf{E}}_{\hat{B}}
    \rangle\!\rangle_{\mathbf{G}}=\delta_{\hat{A}\hat{B}}\,.
\end{equation}
%-----------------------------
The transformation between the two bases is given by a
matrix $\textsf{F}_{\hat{A}}{}^B$ as
%-----------------------------
\begin{equation}\label{orthonormal-basis}
    \hat{\mathbf{E}}_{\hat{A}}=\textsf{F}_{\hat{A}}{}^B~\mathbf{E}_{B}\,.
\end{equation}
%-----------------------------
The orthonormality condition gives
%-----------------------------
\begin{equation}\label{orthonormal-f}
  \textsf{F}_{\hat{A}}{}^{C}\textsf{F}_{\hat{B}}{}^D
  G_{CD}=\delta_{\hat{A}\hat{B}}.
\end{equation}
%-----------------------------
Any $\textsf{F}_{\hat{A}}{}^{C}$ that satisfies this equation
gives an orthonormal basis. Given such an
$\textsf{F}_{\hat{A}}{}^{C}$, we can also obtain an orthonormal
basis for the dual space by using its inverse. Defining
$\textsf{F}^{\hat{C}}{}_D$ as the inverse of the matrix
$\textsf{F}_{\hat{A}}{}^B$, i.e.,  $\textsf{F}_{\hat{A}}{}^B
\textsf{F}^{\hat{A}}{}_C = \delta_C^B$ and
$\textsf{F}_{\hat{A}}{}^B \textsf{F}^{\hat{C}}{}_B =
\delta_{\hat{A}}^{\hat{C}}$, we obtain the dual orthonormal basis
$\{\hat{\mathbf{E}}^{\hat{A}}\}$ in terms of the original dual
basis $\{\mathbf{E}^A\}$ by
%-----------------------------
\begin{equation}
  \hat{\mathbf{E}}^{\hat{A}} = \textsf{F}^{\hat{A}}{}_{B}
  \mathbf{E}^B.
\end{equation}
%-----------------------------

For thermal stresses, assuming that the initial material manifold
is Euclidean, $G_{CD} = e^{2\omega(T)}\delta_{CD} =
\vartheta(T)^{2}\delta_{CD}$ gives
%-----------------------------
\begin{equation}\label{orthonormal-f-for-t}
  \textsf{F}_{\hat{A}}{}^C = \delta_{\hat{A}}^C ~e^{-\omega(T)} =  \delta_{\hat{A}}^C~\vartheta^{-1}(T),
\end{equation}
%-----------------------------
as a solution to (\ref{orthonormal-f}). Here, $\delta_{\hat{A}}^B$ is 1 for
$A=B$, and 0, otherwise, i.e., $\delta_{\hat{1}}^1 =
\delta_{\hat{2}}^2 = \delta_{\hat{3}}^3 = 1$, etc.
Note that (\ref{orthonormal-f}) has
other solutions, too, which we will comment
on below.

Now let us write the components of the total deformation gradient $\mathbf{F}$ in the
orthonormal basis $\{\hat{\mathbf{E}}_{\hat{A}}\}$. The components
are transformed by using $\textsf{F}$:
%-----------------------------
\begin{equation}
  F^{a}{}_{\hat{A}} = \textsf{F}_{\hat{A}}{}^B F^{a}{}_B\,.
\end{equation}
%-----------------------------
Now, using (\ref{orthonormal-f-for-t}),  (\ref{alpha-theta-relation}),
and (\ref{ft-definition}),
we see that
the components $F^a{}_{\hat{A}}$ are given precisely by
those of $\mathbf{F}_e$, the ``elastic part'' of the
deformation gradient in Stojanovitch's approach:
%-----------------------------
\begin{equation}
	F^{a}{}_{\hat{A}} = \textsf{F}_{\hat{A}}{}^B F^{a}{}_B= \delta_{\hat{A}}^B e^{-\omega(T)} F^{a}{}_B 
   = (\vartheta(T))^{-1} \delta_{\hat{A}}^B F^{a}{}_B= (F_T^{-1})_{A}{}^B F^{a}{}_B 
  = (F_e)^a{}_{A}.
\end{equation}
%-----------------------------
Thus, Stojanovi\'c's $\mathbf{F}_e$ is nothing but the original
deformation gradient, written in terms of an orthonormal basis in
the material space. In passing, we have also shown that there is
no need for a mysterious ``intermediate configuration'' as the
target space of $\mathbf{F}_T$, the latter just gives an
orthonormal frame in the material manifold, and as such, can be
treated as a linear map from the tangent space of the material
manifold to itself. As mentioned above, these ideas can be
generalized to other problems with residual stresses.

Rewriting the constitutive relation (\ref{constitutive-riemann})
by using an orthonormal basis for the material manifold, we obtain
%-----------------------------
\begin{equation}
  \Psi=\Psi(\mathbf{X},T,G_{AB} =
  \delta_{AB},F^a{}_{B} = (F_e)^a{}_B ,g_{ab}).
\end{equation}
%-----------------------------
Thus, given a constitutive relation $\Psi^{\textrm{Riem}}$ in the
Riemannian approach, one can obtain a constitutive relation
$\Psi^{\textrm{LR}}$ in the ``local relaxation'' approach by
simply going to an orthonormal basis by (\ref{orthonormal-basis})
and (\ref{orthonormal-f}), and ignoring the constant terms $G_{AB}
= \delta_{AB}$ and $g_{ab} = \delta_{ab}$ in the functional
dependence.
%-----------------------------
\begin{equation}
  \Psi^{\textrm{LR}}(\mathbf{X}, T,(F_e)^a{}_B) =  \Psi^{\textrm{Riem}}\left(\mathbf{X},T,G_{AB} =
  \delta_{AB},F^a{}_{B} = (F_e)^a{}_B ,g_{ab}=\delta_{ab}\right).
\end{equation}
%-----------------------------
Going in the opposite direction is also possible; starting with a
free energy function for the Stojanovi\'c's approach, one can
derive an equivalent free energy in the Riemannian approach. This
direction may be slightly more confusing, since the metrics of the
material manifold and the spatial manifold are not explicitly
written out initially. One proceeds by first writing $\mathbf{F}_e$ in terms of its
proper index structure $(F_e)^a{}_B$ in the Riemannian approach,
and inserting $\delta_{ab}$ and $\delta_{AB}$ where necessary for
tensorial consistency, and finally interpreting these as the
components of metric tensors, and performing a change of basis, if
desired.

\paragraph{Non-coordinate bases and torsion.}
Although a coordinate basis $\{\mathbf{E}_A = \partial / \partial
X^A\}$ is not necessarily orthonormal, one can always obtain an
orthonormal basis by applying a pointwise change of basis
$\mathsf{F}_{\hat{A}}{}^B$. Moreover, giving an orthonormal basis
in this way is equivalent to giving a metric tensor at each point;
the inner product of any two vectors can be calculated by using
their components in the orthonormal basis. We have seen above that
in the context of thermo-elasticity, this means that a change in
the material metric due to a change in temperature can be given in
terms of the ``thermal deformation gradient'' of the local
relaxation approach.

Given an orthonormal basis $\{\hat{\mathbf{E}}_{\hat{A}}\}$, it is
possible to obtain another one, $\{\hat{\mathbf{E}}'_{\hat{A}}\}$,
by using an orthogonal transformation
$\Lambda_{\hat{A}}{}^{\hat{B}}$ as
%-----------------------------
\begin{equation}
   \hat{\mathbf{E}}'_{A}=\Lambda_{\hat{A}}{}^{\hat{B}}~\hat{\mathbf{E}}_{\hat{B}},
\end{equation}
%-----------------------------
where $\Lambda_{\hat{A}}{}^{\hat{B}}$ satisfies
$\Lambda_{\hat{A}}{}^{\hat{C}} \Lambda_{\hat{B}}{}^{\hat{D}}
\delta_{\hat{C}\hat{D}} = \delta_{\hat{A}\hat{B}}$. Let the relation between the original coordinate basis
$\{\mathbf{E}_{A}\}$ and the new orthonormal basis be given by the
matrix ${\textsf{F}'}_{\hat{A}}{}^B$ as follows
%-----------------------------
\begin{equation}
    \hat{\mathbf{E}}'_{\hat{A}}=\textsf{F}'_{\hat{A}}{}^B \mathbf{E}_B.
\end{equation}
%-----------------------------
The relation between $\mathsf{F}$ and $\mathsf{F}'$ is given as
%-----------------------------
\begin{equation}\label{oldf-newf}
    \textsf{F}'_{\hat{A}}{}^B=\Lambda_{\hat{A}}{}^{\hat{C}}\textsf{F}_{\hat{C}}{}^B.
\end{equation}
%-----------------------------
Going in the opposite direction, one can see that $\mathsf{F}$ and
$\mathsf{F}'$ represent the same material metric $\mathbf{G}$, if
and only if they are related through (\ref{oldf-newf}) for some
orthogonal matrix $\Lambda_{\hat{A}}{}^{\hat{B}}$. This means that
there is an $SO(3)$ ambiguity in the choice of $\mathsf{F}$, and
hence, in that of $\mathbf{F}_T$.

As opposed to a coordinate basis, the elements of an orthonormal
basis do not necessarily commute with each other;  whereas
$[\mathbf{E}_A,\mathbf{E}_B]=\mathbf{0}$ for $E_A =
\partial/\partial X^A$, for an orhonormal basis $\hat{\mathbf{E}}_{\hat{A}} =
\textsf{F}_{\hat{A}}{}^B \mathbf{E}_B$, one has
\citep{Nakahara2003}
%-----------------------------
\begin{equation}
    [\hat{\mathbf{E}}_{\hat{A}},\hat{\mathbf{E}}_{\hat{B}}]=c_{\hat{A}\hat{B}}{}^{\hat{C}}~\hat{\mathbf{E}}_{\hat{C}},
\end{equation}
%-----------------------------
where
%-----------------------------
\begin{equation}\label{c-orthonormal}
    c_{\hat{A}\hat{B}}{}^{\hat{C}} = \textsf{F}^{\hat{C}}{}_D\left(
    \textsf{F}_{\hat{A}}{}^E\frac{\partial \textsf{F}_{\hat{B}}{}^D}{\partial X^E}
    -\textsf{F}_{\hat{B}}{}^E\frac{\partial \textsf{F}_{\hat{A}}{}^D}{\partial X^E}\right).
\end{equation}
%-----------------------------
The connection coefficients
$\overline{\Gamma}_{\hat{A}\hat{B}}^{\hat{C}}$ for an orthonormal
basis, defined through
%-----------------------------
\begin{equation}
    \nabla_{\hat{A}}\hat{\mathbf{E}}_{\hat{B}}=\overline{\Gamma}_{\hat{A}\hat{B}}^{\hat{C}} ~\hat{\mathbf{E}}_{\hat{C}},
\end{equation}
%-----------------------------
are related to the connection coefficients $\Gamma_{AB}^C $of the
coordinate basis by
%-----------------------------
\begin{equation}\label{connection-orthonormal}
    \overline{\Gamma}_{\hat{A}\hat{B}}^{\hat{C}} =
      \textsf{F}_{\hat{A}}{}^D\textsf{F}^{\hat{C}}{}_F\left(\frac{\partial \textsf{F}_{\hat{B}}{}^F}{\partial X^D}+\textsf{F}_{\hat{B}}{}^E\Gamma_{DE}^F\right).
\end{equation}
%-----------------------------

In a coordinate basis, the components of the torsion tensor are
given by the antisymmetrization of the two lower indices of the
connection coefficients, i.e. (see the appendix)
%-----------------------------
\begin{equation}\label{torsion-coord}
    T_{AB}{}^{C}=\Gamma_{AB}^{C}-\Gamma_{BA}^{C}.
\end{equation}
%-----------------------------
However, for a non-coordinate basis, the components
are given by
%-----------------------------
\begin{equation}\label{torsion-orthonormal}
    T^{\hat{C}}{}_{\hat{A}\hat{B}}=\overline{\Gamma}_{\hat{A}\hat{B}}^{\hat{C}}-\overline{\Gamma}_{\hat{B}\hat{A}}^{\hat{C}}-c_{\hat{A}\hat{B}}{}^{\hat{C}}.
\end{equation}
%-----------------------------
Our formalism is based on Riemannian geometry, and in particular,
on the torsion-free Levi-Civita connection defined by the metric
$\mathbf{G}(T)$. Thus, the torsion tensor for the material
connection vanishes in both the coordinate basis, and the
orthonormal basis. Let us show this explicitly for
$\mathsf{F}_{\hat{A}}{}^B = \vartheta^{-1}(T) \delta_{\hat{A}}^B$.
In the original coordinate basis, the metric tensor is given by
$G_{AB}=e^{2\omega(T)}\delta_{AB} = \vartheta^2(T)\delta_{AB}$.
Thus, the connection coefficients in this basis are
%-----------------------------
\begin{equation}
    \Gamma_{BC}^A=\vartheta^{-1}\left(\vartheta_{,B}\delta^A_C+\vartheta_{,C}\delta^A_B-\vartheta_{,D}\delta^{AD}\delta_{BC}\right).
\end{equation}
%-----------------------------
Using (\ref{torsion-coord}), we have $T^A{}_{BC}=0$. Next, using (\ref{c-orthonormal}) and (\ref{connection-orthonormal})
with
%-----------------------------
\begin{equation}
    \textsf{F}^{\hat{A}}{}_B=\vartheta~ \delta^{\hat{A}}_B,~~~\textsf{F}_{\hat{A}}{}^B=\vartheta^{-1} \delta_{\hat{A}}^B,
\end{equation}
%-----------------------------
we obtain
%-----------------------------
\begin{equation}
    c_{\hat{A}\hat{B}}{}^{\hat{C}}=\frac{\vartheta_{,D}}{\vartheta^2}\left(\delta_{\hat{A}}^{\hat{C}}\delta_{\hat{B}}^D-\delta_{\hat{B}}^{\hat{C}}\delta_{\hat{A}}^D\right),
\end{equation}
%-----------------------------
and
%-----------------------------
\begin{equation}
    \overline{\Gamma}_{\hat{A}\hat{B}}^{\hat{C}}=-\frac{\vartheta_{,D}}{\vartheta^2}\delta_{\hat{A}}^D\delta_{\hat{B}}^{\hat{C}}
    +\frac{1}{\vartheta}\delta_{\hat{A}}^D\delta_{\hat{B}}^E\delta_C^{\hat{C}}\Gamma_{DE}{}^C.
\end{equation}
%-----------------------------
Using these in (\ref{torsion-orthonormal}), we obtain
$\overline{T}_{\hat{B}\hat{C}}{}^{\hat{A}}=0$. Note that the
Riemann curvature tensor has the following form
%-----------------------------
\begin{eqnarray}
 \mathcal{R}^A{}_{BCD}&=&\vartheta^{-2}\left[2\left(\vartheta_{,C}\delta^A_B-\vartheta_{,B}\delta^A_C\right)\vartheta_{,D}
  +2\left(\vartheta_{,B}\delta_{CD}-\vartheta_{,C}\delta_{BD}\right)\vartheta_{,E}\delta^{AE}
  +\vartheta_{,E}\vartheta_{,F}\delta^{EF}\left(\delta^A_C\delta_{BD}-\delta^A_B\delta_{CD}\right)\right]   \nonumber \\
  &&
  +\vartheta^{-1}\left[\vartheta_{,BD}\delta^A_C-\vartheta_{,CD}\delta^A_B+\left(\vartheta_{,CE}\delta_{BD}-\vartheta_{,BD}\delta_{CD}\right)\delta^{AE}\right],
\end{eqnarray}
%-----------------------------
which, in general, does not vanish.

Stojanovi\'c, on the other hand, calculates a non-vanishing
torsion tensor. This discrepancy is due to the fact that he uses
the following connection
%-----------------------------
\begin{equation}
    \Gamma^A_{BC}=\left(F_T^{-1}\right)^A{}_{M}\frac{\partial \left(F_T\right)^M{}_C}{\partial X^B}.
\end{equation}
%-----------------------------
Similar connections have been used in other contexts
\citep{Kondo1955a,BilbyBulloughSmith1955}. It can be shown that
this connection has vanishing curvature but has nonvanishing
torsion. This is related to the so-called canonical connection in absolutely parallelizable manifolds
\citep{Eisenhart1926,Eisenhart1927,YoussefSid-Ahmed2007}. See also
\citet{EpsteinElzanowski2007} for similar connections in the
context of inhomogeneities and their geometric representations. In
the appendix, we give some details on absolutely parallelizable
manifolds and the above connection.

In summary, our geometric approach has a concrete connection with
that of Stojanovi\'c: in our approach we use a Riemannian manifold
with a temperature-dependent metric as the material manifold while
Stojanovi\'c implicitly uses the same metric but in an absolutely
parallelizable manifold that is not Riemannian. Using either
approach would be fine and a matter of taste, however we believe
that our approach is more straightforward as we do not introduce
an unnecessary torsion in the material manifold. Representing
changes of temperature by a one-parameter family of conformal
Riemannian metrics enables us to find the zero-stress temperature
distributions even in finite deformations.

\subsection{Anisotropic Thermal Expansion}

So far, we have assumed that thermal expansion is isotropic, i.e.
a change of temperature results in a change of length independent
of orientation. Let us now see how one should modify the theory
when a temperature change results in different changes in length
in different directions and in possibly a change of shape. Even in
the case of anisotropic thermal expansion all one needs is a
temperature-dependent material metric $\mathbf{G}(\mathbf{X},T)$
but in this case the material metric is no longer a simple
rescaling of the original material metric. The physical idea is
the following. Given any point in the initial stress-free material
manifold with metric $\mathbf{G}_0(\mathbf{X})$, there exists a
frame field
$\{\mathbf{E}^1(\mathbf{X}),\mathbf{E}^2(\mathbf{X}),\mathbf{E}^3(\mathbf{X})\}$
for each material point such that in this frame the material
metric is diagonal and has the following form
%-----------------------------
\begin{equation}
    \mathbf{G}(\mathbf{X},T)=e^{2\omega_1(T)}\mathbf{E}^1\otimes\mathbf{E}^1+e^{2\omega_2(T)}\mathbf{E}^2\otimes\mathbf{E}^2+e^{2\omega_3(T)}\mathbf{E}^3\otimes\mathbf{E}^3.
\end{equation}
%-----------------------------
Given a coordinate basis
$\widetilde{\mathbf{E}}_I=\partial/\partial X^I$, we have
%-----------------------------
\begin{equation}
    \mathbf{E}_I=A^J{}_I\widetilde{\mathbf{E}}_J.
\end{equation}
%-----------------------------
Thus
%-----------------------------
\begin{equation}
    \mathbf{G}(\mathbf{X},T)=\sum_{I}e^{2\omega_I(T)}A^J{}_I\widetilde{\mathbf{E}}_J\otimes A^K{}_I\widetilde{\mathbf{E}}_K.
\end{equation}
%-----------------------------

\section{Geometric Elasticity with Temperature Changes}

\paragraph{Material manifold.} As mentioned in the previous sections,
the material manifold describes the intrinsic ``shape'' of the
natural, stress-free state of the material. The geometry induced
by a given configuration of the material in the spatial manifold
may or may not agree with the intrinsic geometry. The discrepancy
between the two geometries (the induced and the intrinsic) is in
general a cause for stresses, which are described by geometric
constitutive relations. In this section, we will describe this
framework.\footnote{Due to our approach to thermal stresses, in
this paper, we treat the material and the spatial spaces as
Riemannian manifolds, as in \citep{MaHu1983,YaMaOr2006}, and by
``geometry'', we understand the Levi-Civita connection associated
to the metric tensor. In general, the geometry of either of these
spaces can be given by a more general connection that has torsion
and/or non-metricity. Such connections have found use in the
literature of defect mechanics.}

\begin{figure}[hbt]
\begin{center}
\includegraphics[scale=0.65,angle=0]{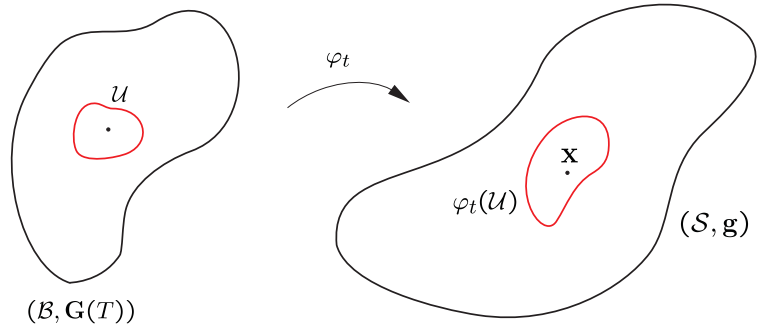}
\end{center}
\caption{\footnotesize Motion of a continuum with temperature
changes.} \label{Deformation}
\end{figure}

The motion of an elastic body is described by a possibly
non-isometric, time dependent embedding of the material manifold
in the spatial manifold (see Fig. \ref{Deformation}). There is, however, another possible
source of time-dependency: the geometry of the material manifold
itself may change in time.\footnote{The geometry of the spatial
manifold may also change, but we do not consider this issue in
this paper.}  A change in the geometry of the material manifold is
sometimes known as a referential change (see \citep{Gur00,Mau93}
and references therein), and the precise meaning of this has
sometimes been a source of confusion in the literature. In this
paper, we have a case where the change in the geometry of the
material manifold is described explicitly in terms of the
temperature, and we believe that the conceptual clarity brought by
this simple example may provide insights to other cases of
referential changes, such as those that describe the evolution of
defects in a crystalline solid.

\vskip 0.1 in
%-----------------------------
%-----------------------------
%-----------------------------
%-----------------------------

In Section 2, we proposed to describe the thermal expansion of an
isotropic material by a change in the material metric, given by
equation (\ref{material-metric}):
%-----------------------------
\begin{equation}\label{material-metric-repeat}
  G_{IJ}(\mathbf{X},T) = H_{IJ}(\mathbf{X}) e^{2 \omega(T)}\,.
\end{equation}
%-----------------------------
Here, $\omega(T)$ is a function that describes the thermal expansion
properties of the isotropic material under consideration.  The
coefficient of thermal expansion is given by equation
(\ref{coefficient-thermal-expansion}):
%-----------------------------
\begin{equation}
  \alpha(T) = \frac{d\omega(T)}{dT}\,.
\end{equation}
%-----------------------------
We assume that the temperature distribution in the material is
given. If the temperature depends on time, then the material
metric describing the relaxed state of the material will also
depend on time, through (\ref{material-metric-repeat}).

We should mention that evolution of reference configuration in the
literature of continuum mechanics is more or less ambiguous. It is
believed that an evolving reference configuration can model
dynamics of defects. However, to our best knowledge, there are no
concrete examples in the literature. We believe that the present
geometric formulation of thermal stresses in terms of a
temperature-dependent material manifold can make the role of
reference manifold clearer and can shed light on other more
complicated problems, e.g., continuum theory of solids with
distributed dislocations.

\paragraph{Conservation of mass.} Let us begin by writing the
conservation of mass in this setting. If the temperature is time-independent,
the usual material version of the conservation of mass holds: the
material density is constant.
%-----------------------------
\begin{equation}
    \rho_0(\mathbf{X},t)= \rho_0(\mathbf{X}).
\end{equation}
%-----------------------------
If, however, temperature changes in time, the material metric will
expand or contract, so the material mass density will change. The
evolution of the mass density will then be given by
%-----------------------------
\begin{equation}\label{material-density}
    \rho_0(\mathbf{X},T)dV(\mathbf{X},T)=\textsf{m}(\mathbf{X}),
\end{equation}
%-----------------------------
where $dV(\mathbf{X},T(t))$ is the volume form of the metric
$G(\mathbf{X},T)$, and $\textsf{m}(\mathbf{X})$ is the
temperature-independent (and hence time-independent) differential
form representing the mass density (mass form). This equation
tells us that if the material manifold expands due to a
temperature change, the total mass in a material region will not
change, and hence the density $\rho_0$ will decrease inversely with
the increase in the volume of that region. Since the volume form
is given by
%-----------------------------
\begin{equation}
  dV(\mathbf{X},T) = \sqrt{\det{|H_{IJ}|}} ~e^{N \omega(T)} d^NX,
\end{equation}
%-----------------------------
we can get the density for a given temperature $T$ in terms of the
density at an initial temperature $T_0$ by using
$\rho_0(\mathbf{X},T)dV(\mathbf{X},T) =
\rho_0(\mathbf{X},T_0)dV(\mathbf{X},T_0)$. This gives
%-----------------------------
\begin{equation}
  \rho_0(\mathbf{X}, T) = e^{N(\omega(T_0)-\omega(T))}
  \rho_0(\mathbf{X},T_0).
\end{equation}
%-----------------------------
In terms of the coefficient of thermal expansion $\alpha =
\frac{d\omega}{dT}$, this can be written as
%-----------------------------
\begin{equation}
    \rho_0(\mathbf{X},T)=\rho_0(\mathbf{X},T_0)e^{-N\int_{T_0}^{T}\alpha(\tau)d\tau}.
\end{equation}
%-----------------------------

\paragraph{Incompressibility.} Elastic incompressibility means that
elastic deformations cannot cause any changes in volume. Thus, for
a given temperature distribution, the deformation map must
preserve the volume element. The volume elements in the material
and spatial manifolds, $dV(\mathbf{X})$ and $dv(\mathbf{x})$ are
related by
%-----------------------------
\begin{equation}
    dv(\mathbf{x(\mathbf{X})})=J(\mathbf{X},T)dV(\mathbf{X},T),
\end{equation}
%-----------------------------
where the Jacobian $J$ is given as
%-----------------------------
\begin{equation}
    J(\mathbf{X},T)=\det \mathbf{F} \sqrt{\frac{\det \mathbf{g}}{\det
    \mathbf{G}(T)}}.
\end{equation}
%-----------------------------
Thus, incompressibility means that $J(\mathbf{X},T)=1$. Given two
Riemannian manifolds, distance preserving maps, i.e., isometries
between them may or may not exist. A similar question may arise
for volume-preserving maps: given two Riemannian manifolds (in our
case, the material manifold and the spatial manifold), does there
exist a volume-preserving map between them? \citet{Moser1965}
answers this question in the affirmative, so the study of
incompressibility in this setting is not vacuous.

\paragraph{Free energy.} The free energy, in addition to explicitly depending on temperature, will depend on the
temperature-dependent material metric tensor as well, i.e.
%-----------------------------
\begin{equation}
    \Psi=\Psi(\mathbf{X},T,\mathbf{G}(\mathbf{X},T),\mathbf{F},\mathbf{g}).
\end{equation}
%-----------------------------
Therefor, the first Piola-Kirchoff stress, given by
%-----------------------------
\begin{equation}
    \mathbf{P}=\mathbf{P}(\mathbf{X},T)=\mathbf{g}^{-1}\frac{\partial\Psi}{\partial \mathbf{F}}\,,
\end{equation}
%-----------------------------
explicitly depends on the temperature-dependent material metric.

\paragraph{Balance of Linear Momentum.} Let us now look at the governing
equations for a given temperature distribution $T=T(\mathbf{X})$.
We will only study the static case\footnote{The governing
equations for the dynamics case are similar. However, for the
dynamic problem one has to consider an evolving temperature
distribution governed by the heat equation. This will be discussed
in a future communication.}, for which the balance of linear
momentum reads
%-----------------------------
\begin{equation}\label{elastostatics}
    \operatorname{Div}\mathbf{P}=\mathbf{0}~~~~~\textrm{or}~~~~~P^{aA}{}_{|A}=\frac{\partial
    P^{aA}}{\partial X^A}+\Gamma_{AB}^A P^{aB} + \gamma_{bc}^a F^c{}_A
    P^{bA}=0,
\end{equation}
%-----------------------------
where $\Gamma_{AB}^C$ are the connection coefficients for the
material metric $G_{AB}$, and $\gamma_{ab}^c$ are the connection
coefficients for the metric $g_{ab}$.  This is the standard
balance of momentum in geometric elastostatics, see, e.g.,
\citet{YaMaOr2006}. For the case of thermal stresses, the material
connection coefficients $\Gamma_{AB}^C$ are those of the metric
(\ref{material-metric}), $G_{IJ}(\mathbf{X},T) =
H_{IJ}(\mathbf{X}) e^{2 \omega(T)}$; they are given in terms of
the connection coefficients $\Gamma_{AB}^{(H)~C}$ of the metric
$H_{IJ}$ as \citep{Wald1984}
%-----------------------------
\begin{equation}
  \Gamma_{AB}^C =
  \Gamma_{AB}^{(H)\,C}+\left(\delta^C_{A}\partial_{B}\Omega + \delta^C_{B}\partial_{A}\Omega -
  H_{AB}H^{CD}\partial_D\Omega\right).
\end{equation}
%-----------------------------
Suppose the initial material metric $H_{AB}$ is Euclidean. Then,
using Cartesian coordinates, we have
%-----------------------------
\begin{equation}
  \Gamma_{AB}^A = 3\partial_B\Omega.
\end{equation}
%-----------------------------
For a Euclidean spatial metric in Cartesian coordinates,
$g_{ab}=\delta_{ab}$, we obtain
%-----------------------------
\begin{equation}\label{linear-momentum-temperature}
    \frac{\partial P^{aA}}{\partial X^A}+3 \frac{\partial \Omega}{\partial X^B} P^{aB}=0.
\end{equation}
%-----------------------------
In terms of the thermal expansion coefficient
$\alpha=\frac{d\omega}{dT}$, this becomes
%-----------------------------
\begin{equation}\label{linear-momentum-temperature}
    \frac{\partial P^{aA}}{\partial X^A}+3\alpha \frac{\partial T}{\partial X^B} P^{aB}=0.
\end{equation}
%-----------------------------
In the following example, we show that in the geometric framework,
some nonlinear problems can be solved analytically.

\paragraph{Example.} Let us consider a two-dimensional, incompressible
neo-Hookean material in a flat two-dimensional spatial manifold. The
free energy density of a neo-Hookean material in two dimensions has the form
%-----------------------------
\begin{equation}
    \Psi=\Psi(\mathbf{X},\mathbf{C})=\mu(\operatorname{tr}\mathbf{C}-2),
\end{equation}
%-----------------------------
where $\mathbf{C}$ is the Cauchy-Green tensor, or equivalently,
the pull-back of the spatial metric, $C_{AB} = F^a{}_AF^b{}_B
g_{ab}$, and $\mu$ is a material constant. We will assume that
this form holds for an isotropic material under thermal expansion,
and in particular, we assume that there is no explicit temperature
dependence in the free energy apart from the dependence through
$\mathbf{C}$. In components
%-----------------------------
\begin{equation}
    \Psi=\mu\left(F^a{}_AF^b{}_Bg_{ab}G^{AB}-2\right).
\end{equation}
%-----------------------------
The ``2'' is of no particular significance: when the material
metric is fixed, it just shifts the free energy by a constant.
When the material metric changes as in (\ref{material-metric}),
its contribution to the free energy is proportional to the
temperature-dependent material volume, which, for a given
temperature distribution, is independent of the spatial
configuration. We ignore this term, and use
$\Psi=\Psi(\mathbf{X},\mathbf{C})=\mu\operatorname{tr}\mathbf{C}$
as our definition of the free energy.

Let us assume that initially the material has a flat annular shape
$R_1\le R \le R_2$ without any stresses, at a uniform temperature
$T_0$. We would like to calculate the stresses that occur in the
new equilibrium configuration after we change the temperature in a
rotationally symmetric way, $T=T(R)$. In polar coordinates, the spatial metric and its inverse read
%-----------------------------
\begin{equation}
    \mathbf{g}=\left(%
\begin{array}{cc}
  g_{rr} & g_{r\theta} \\
  g_{\theta r} & g_{\theta\theta} \\
\end{array}%
\right)=\left(%
\begin{array}{cc}
  1 & 0 \\
  0 & r^2 \\
\end{array}%
\right),~~~
\mathbf{g}^{-1}=\left(%
\begin{array}{cc}
  g^{rr} & g^{r\theta} \\
  g^{\theta r} & g^{\theta\theta} \\
\end{array}%
\right)=\left(%
\begin{array}{cc}
  1 & 0 \\
  0 & 1/r^2 \\
\end{array}%
\right),
\end{equation}
%-----------------------------
and thus $\det \mathbf{g}=r^2$. The only nonzero connection
coefficients are:
%-----------------------------
\begin{equation}
    \gamma_{\theta\theta}^r=-r,~\gamma_{r\theta}^{\theta}=\gamma_{\theta r}^{\theta}=1/r.
\end{equation}
%-----------------------------
For the temperature-dependent material metric we have
%-----------------------------
\begin{equation}
    \mathbf{G}=\left(%
\begin{array}{cc}
  G_{RR} & G_{R\Theta} \\
  G_{\Theta R} & G_{\Theta\Theta} \\
\end{array}%
\right)=\left(%
\begin{array}{cc}
  1 & 0 \\
  0 & R^2 \\
\end{array}%
\right)e^{2\omega(T(R))},~~~
\mathbf{G}^{-1}=\left(%
\begin{array}{cc}
  G^{RR} & G^{R\Theta} \\
  G^{\Theta R} & G^{\Theta\Theta} \\
\end{array}%
\right)=\left(%
\begin{array}{cc}
  1 & 0 \\
  0 & 1/R^2 \\
\end{array}%
\right)e^{-2\omega(T(R))},
\end{equation}
%-----------------------------
and thus, $\det \mathbf{G}=R^2e^{4\omega(T(R))}$. The following
nonzero connection coefficients are needed in the balance of
linear momentum:
%-----------------------------
\begin{equation}
    \Gamma_{RR}^R=\Omega'(R),~\Gamma_{\Theta\Theta}^R=-R-R^2 \Omega'(R),~\Gamma_{R\Theta}^{\Theta}=\Gamma_{\Theta R}^{\Theta}=1/R+\Omega'(R).
\end{equation}
%-----------------------------
In terms of the thermal expansion coefficient, these are given as
%-----------------------------
\begin{equation}
    \Gamma_{RR}^R=\alpha T'(R),~\Gamma_{\Theta\Theta}^R=-R-R^2 \alpha T'(R),~\Gamma_{R\Theta}^{\Theta}=\Gamma_{\Theta R}^{\Theta}=1/R+\alpha T'(R).
\end{equation}
%-----------------------------
Given the temperature distribution $T=T(R)$, we are looking for
solutions of the form
%-----------------------------
\begin{equation}
  \varphi(R,\Theta)=(r,\theta)=(r(R),\Theta).
\end{equation}
%-----------------------------
Thus
%-----------------------------
\begin{equation}
    \mathbf{F}=\left(%
\begin{array}{cc}
  r'(R) & 0 \\
  0 & 1 \\
\end{array}%
\right),~~~\mathbf{F}^{-1}=\left(%
\begin{array}{cc}
  1/r'(R) & 0 \\
  0 & 1 \\
\end{array}%
\right).
\end{equation}
%-----------------------------
This gives the Jacobian as
%-----------------------------
\begin{equation}
    J=\frac{r\,r'}{Re^{2\omega(T)}}.
\end{equation}
%-----------------------------
Incompressibility dictates that
%-----------------------------
\begin{equation}\label{incomp}
    rr'=R e^{2\omega(T)}.
\end{equation}
%-----------------------------
This differential equation has the following solution
%-----------------------------
\begin{equation}
    r^2(R)=\int_{R_1}^{R}2\xi e^{2\omega(T(\xi))} d\xi+r_1^2(R).
\end{equation}
%-----------------------------
Note that $r_1(R)$ is not known a priori and will be obtained
after imposing the traction boundary conditions at $r_1$ and
$r_2$.

In incompressible elasticity, $P^{aA}$ is replaced by
$P^{aA}-Jp(F^{-1})^{-A}{}_bg^{ab}$, where $p$ is an unknown scalar field
(pressure) that will be determined using the constraint $J=1$
\citep{MaHu1983}, i.e.
%-----------------------------
\begin{equation}
    P^{aA}=2\mu F^a{}_BG^{AB}-p(R)(F^{-1})^{A}{}_bg^{ab}.
\end{equation}
%-----------------------------
Therefore, using (\ref{incomp}), we get the nonzero stress components as
%-----------------------------
\begin{equation}
    P^{rR}= \frac{2\mu R}{r}-p(R)\frac{r}{R}e^{-2\omega(T(R))},~~~
    P^{\theta\Theta}=\frac{2\mu}{R^2}e^{-2\omega(T(R))}-\frac{p(R)}{r^2},
\end{equation}
%-----------------------------
where $p(R)$ is an unknown pressure.

Balance of linear momentum in components reads
%-----------------------------
\begin{equation}\label{bal-mom}
    P^{aA}{}|{_A}=\frac{\partial P^{aA}}{\partial X^A}+\Gamma^A_{AB}P^{aB}+P^{bA}\gamma^a_{bc}F^c{}_A=0.
\end{equation}
For the radial direction, $a=r$, we have
%-----------------------------
\begin{eqnarray}
  && P^{rA}{}|{_A}=\frac{\partial P^{rA}}{\partial X^A}+\Gamma^A_{AB}P^{rB}+P^{bA}\gamma^r_{bc}F^c{}_A   \nonumber \\
  && ~~~~~~~~= \frac{\partial P^{rR}}{\partial R}+\left(\Gamma^R_{RR}+\Gamma^{\Theta}_{\Theta
  R}\right)P^{rR}+P^{\theta\Theta}\gamma^{r}_{\theta\theta}F^{\theta}{}_{\Theta}
  \nonumber\\
  && ~~~~~~~~= \frac{\partial P^{rR}}{\partial R}+\left(\frac{1}{R}+2\alpha T'(R)\right)P^{rR}-r P^{\theta\Theta}=0.
\end{eqnarray}
%-----------------------------
This gives
%-----------------------------
\begin{equation}
    p'(R)=\frac{2\mu R}{r^2}e^{2 \omega(T(R))}\left[2\left(1+\alpha R T'\right)-\frac{R^2}{r^2}e^{2\omega(T(R))}-\frac{r^2}{R^2}e^{-2\omega(T(R))}\right].
\end{equation}
%-----------------------------
Assuming that $p(R_i)=0$, we obtain
%-----------------------------
\begin{equation}
    p(R)=\int_{R_i}^{R}\frac{2\mu \xi}{r^2(\xi)}e^{2\omega(T(\xi))}\left[2\left(1+\alpha(\xi) \xi T'(\xi)\right)-\frac{\xi^2}{r^2(\xi)}e^{2\omega(T(\xi))}
    -\frac{r^2(\xi)}{\xi^2}e^{-2\omega(T(\xi))}\right]d\xi.
\end{equation}
%-----------------------------
For $a=\theta$, balance of momentum (\ref{bal-mom}) gives,
%-----------------------------
\begin{equation}
    P^{\theta A}{}|{_A}= \frac{\partial P^{\theta\Theta}}{\partial \Theta}
     +\Gamma^A_{A\Theta}P^{\theta\Theta}+P^{\theta R}\gamma^{\theta}_{rr}F^{r}{}_{R}
      +P^{\theta\Theta}\gamma^{\theta}_{\theta\theta}F^{\theta}{}_{\Theta}
      = \left(\Gamma^R_{R\Theta}+\Gamma^{\Theta}_{\Theta\Theta}\right)P^{\theta\Theta}=0.
\end{equation}
%-----------------------------
i.e. this equilibrium equation is trivially satisfied. Therefore,
given the temperature distribution $T(R)$, we can calculate all
the thermal stresses analytically.

%\vskip 0.2 in
\section{Linearized Theory of Thermal Stresses}

In this section, we linearize the governing equations of the
nonlinear theory presented in the previous section about a
reference motion. Geometric linearization of elasticity was first
introduced by \citet{MaHu1983} and was further developed by
\citet{YavariOzakin2008}. See also \citep{MazzucatoRachele2006}
for similar discussions. Here, we start with a
temperature-dependent material manifold and its motion in an
ambient space. Given a reference motion, we are interested in
obtaining the linearized governing equations with respect to this
motion.  We will assume that the ambient space manifold is
Euclidean. This is not a necessary assumption but it provides a
natural setting for most practical problems of interest and will
simplify the subsequent calculations. For simplicity, we will
restrict attention to time-independent solutions.

\begin{figure}[hbt]
\begin{center}
\includegraphics[scale=0.55,angle=0]{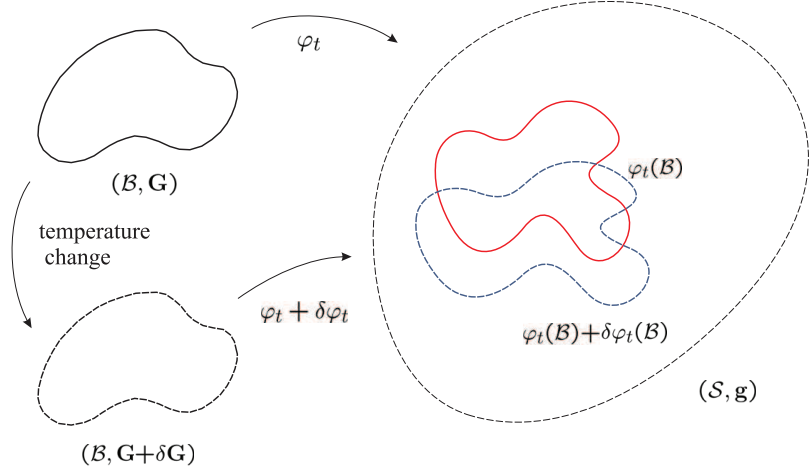}
\end{center}
\caption{\footnotesize Motion of a continuum with temperature
changes.} \label{Deformation_Change}
\end{figure}

Suppose a given material with a temperature distribution
$T(\mathbf{X})$ and the related material metric $\mathbf{G}$ is in
a static equilibrium configuration,  $\varphi$. The balance of
linear momentum for this material body reads
%-----------------------------
\begin{equation}\label{linear-momentum}
    \operatorname{Div}\mathbf{P}+\rho_0\mathbf{B}=\mathbf{0}\,.
\end{equation}
%-----------------------------
Now suppose we change the temperature of this material by a small
amount $\delta T(\mathbf{X})$. This will change the material
metric to $\mathbf{G}' = \mathbf{G}+\delta \mathbf{G}$, and
$\varphi$ will no longer describe a static equilibrium
configuration. A nearby equilibrium configuration may be given by
$\varphi' = \varphi + \delta \varphi$, and the stress in this new
equilibrium configuration will be $\mathbf{P}' =
\mathbf{P}+\delta\mathbf{P}$. One would like to calculate the
change in the stress (or the configuration), for a given small
change in temperature (see Fig. \ref{Deformation_Change}).

\vskip 0.1 in
%-----------------------------
%-----------------------------
%-----------------------------
%-----------------------------

While the spirit of this setup is familiar from other
linearization problems, some care is needed in interpreting its
meaning. $\mathbf{P}(\mathbf{X})$ is a two-point tensor (it has
components in both the material and the ambient spaces: $P^{aA}$)
based at $\mathbf{X}$ and $\varphi(\mathbf{X})$, whereas
$\mathbf{P'}(\mathbf{X})$ is based at $\mathbf{X}$ and
$\varphi'(\mathbf{X})$. Defining $\delta \mathbf{P}(\mathbf{X}) =
\mathbf{P}'-\mathbf{P}$ is nontrivial for a general ambient space
metric. This is related to the fact that subtracting tangent
vectors at different points in a manifold is only defined with
respect to a choice of a path connecting the two points, on which a
parallel transport is to be performed. By restricting our
attention to a Euclidean ambient space we sidestep this issue,
using the natural, path-independent parallel transport in
Euclidean space. Another issue is the definition of $\delta
\varphi = \varphi'-\varphi$. While one can use coordinate systems
to make approximate sense of this equation for two nearby maps, it
is a little troublesome geometrically, since the subtraction of
two maps between manifolds is not defined geometrically.

The linearization procedure can be put to firmer footing if instead of
talking about two nearby configurations and the differences of various
quantities for these configurations, we describe the situation in
terms of a 1-parameter family of configurations around a reference
configuration, and calculate the derivatives of various quantities
with respect to the parameter. These derivatives will capture the
behavior of the solution as a function of the parameter, for small
values of the latter. Thus, let $T_{\epsilon}(\mathbf{X})$ be a 1-parameter family of
temperature distributions on our material manifold,
$\mathbf{G}_{\epsilon}$ be the corresponding family of material metrics,
$\varphi_{\epsilon}$ be the equilibrium configurations, and
$\mathbf{P}_{\epsilon}$ be the stresses. Let $\epsilon=0$ describe
the reference equilibrium configuration. Now, for a fixed point $\mathbf{X}$
in the material manifold, $\varphi_{\epsilon}(\mathbf{X})$
describes a curve in the spatial manifold, and its derivative at
$\epsilon=0$ gives a vector $\mathbf{U(X)}$ at
$\varphi(\mathbf{X})$:
%-----------------------------
\begin{equation}
  \mathbf{U(X)} = \frac{d\varphi_{\epsilon}(\mathbf{X})}{d
  \epsilon}\Big|_{\epsilon=0}.
\end{equation}
%-----------------------------
Considering $\delta \varphi \approx \epsilon
\frac{d\varphi_{\epsilon}}{d \epsilon}$, we see that a more
rigorous version of $\delta \varphi$ is the vector field
$\mathbf{U}$. Similarly, one has
%-----------------------------
\begin{equation}
  \delta\mathbf{G}\approx \epsilon \, \frac{d}{d
  \epsilon}\Big|_{\epsilon=0}\mathbf{G}_{\epsilon}.
\end{equation}
%-----------------------------
When the change in $\mathbf{G}$ is due to a change in $T$, we have
%-----------------------------
\begin{equation}\label{deriv-of-g}
 \frac{d}{d
  \epsilon}\Big|_{\epsilon=0}\mathbf{G}_{\epsilon}=2\frac{d\omega}{dT}\frac{dT}{\partial\epsilon}\Big|_{\epsilon=0}\mathbf{G}
  = \beta \mathbf{G},
\end{equation}
%-----------------------------
where $\beta = 2\frac{d\omega}{dT}\frac{d
T}{d\epsilon}\big|_{\epsilon=0} = 2\alpha(T_{\epsilon})\frac{d
T}{d\epsilon}\big|_{\epsilon=0}$.

Now consider, in the absence of body forces, the equilibrium equation
$\operatorname{Div}\mathbf{P}=\mathbf{0}$ for the family of
temperature distributions parametrized by $\epsilon$:
%-----------------------------
\begin{equation}\label{perturbed-linear-momentum}
    \operatorname{Div}_{\epsilon}\mathbf{P}_{\epsilon}=\mathbf{0}.
\end{equation}
%-----------------------------
Linearization of (\ref{perturbed-linear-momentum}) is defined as
\citep{MaHu1983,YavariOzakin2008}:
%-----------------------------
\begin{equation}\label{perturbed-linear-momentum1}
    \frac{d}{d\epsilon}\Big|_{\epsilon=0}\left(\operatorname{Div}_{\epsilon}\mathbf{P}_{\epsilon}\right)=\mathbf{0}.
\end{equation}
%-----------------------------
Once again, one should note that since the equilibrium
configuration is different for each $\epsilon$,
$\mathbf{P}_{\epsilon}$ is based at different points in the
ambient space for different values of $\epsilon$, and in order to
calculate the derivative with respect to $\epsilon$, one in
general needs to use the connection (parallel transport) in the
ambient space. For the Euclidean case we are considering and a
Cartesian coordinate system $x^a$, this is trivial.  In components
(\ref{perturbed-linear-momentum1}) reads
%-----------------------------
\begin{equation}
    \frac{\partial P^{aA}(\epsilon)}{\partial X^A}+\Gamma^A_{AB}(\epsilon)P^{aB}(\epsilon)=0.
\end{equation}
%-----------------------------
Thus, the linearized balance of linear momentum reads
%-----------------------------
\begin{equation}\label{linearized-balance-mom}
    \frac{\partial}{\partial X^A}\frac{d}{d\epsilon}\Big|_{\epsilon=0}P^{aA}(\epsilon)
    +\left[\frac{d}{d\epsilon}\Big|_{\epsilon=0}\Gamma^A_{AB}(\epsilon)\right]P^{aB}+
    \Gamma^A_{AB}\frac{d}{d\epsilon}\Big|_{\epsilon=0}P^{aB}(\epsilon)=0.
\end{equation}
%-----------------------------
Note that
%-----------------------------
\begin{equation}
  F^{aA} = g^{ac}\frac{\partial\Psi}{\partial F^c{}_A}\,,
\end{equation}
%-----------------------------
where $\Psi=\Psi(\mathbf{X},T,\mathbf{F},\mathbf{G},\mathbf{g})$ is
the material free energy density. In calculating
$\frac{dP^{aA}(\epsilon)}{d\epsilon}$, we need to consider the
changes in $\mathbf{F}$ and $\mathbf{G}$ due to the change in the
equilibrium configuration:
%-----------------------------
\begin{equation}
  \frac{dP^{aA}(\epsilon)}{d\epsilon} = \frac{\partial
  P^{aA}}{\partial F^{b}{}_B}\frac{d F^b{}_B}{d\epsilon} +  \frac{\partial
    P^{aA}}{\partial G_{CD}}\frac{d G_{CD}}{d\epsilon}.
\end{equation}
%-----------------------------
Defining
%-----------------------------
\begin{equation}
  \mathbbm{A}^{aA}{}_b{}^B = \frac{\partial P^{aA}}{\partial F^b{}_B}
  =
  g^{ac}\frac{\partial^2 \Psi}{\partial F^b{}_B \partial F^c{}_A}
  ~~~~~\textrm{and}~~~~~
  \mathbbm{B}^{aACD} = \frac{P^{aA}}{G_{CD}} =
  \frac{\partial^2 \Psi}{\partial G_{CD} \partial F^c{}_A},
\end{equation}
%-----------------------------
where the derivatives are to be evaluated at the reference
configuration $\epsilon = 0$. Noting that
%-----------------------------
\begin{equation}
  \frac{d F^a{}_A}{d\epsilon}\Big|_{\epsilon=0} = \frac{\partial U^a}{\partial
  X^{A}},
\end{equation}
%-----------------------------
we obtain
%-----------------------------
\begin{equation}
    \frac{d}{d\epsilon}\Big|_{\epsilon=0}P^{aA}(\epsilon)
    =\mathbbm{A}^{aA}{}_b{}^BU^b{}_{,B}+\mathbbm{B}^{aACD}G_{CD}~
    \beta.
\end{equation}
%-----------------------------
Using
%-----------------------------
\begin{equation}
    \Gamma^A_{BC}=\frac{1}{2}G^{AD}\left(\frac{\partial G_{BD}}{\partial X^C}+
    \frac{\partial G_{CD}}{\partial X^B}-\frac{\partial G_{BC}}{\partial
    X^D}\right),
\end{equation}
%-----------------------------
and
%-----------------------------
\begin{equation}
  \frac{d G^{AB}}{d \epsilon} =
  -G^{AC}G^{BD}\frac{\partial G_{BD}}{\partial\epsilon},
\end{equation}
%-----------------------------
and plugging in (\ref{deriv-of-g}), we obtain
%-----------------------------
\begin{equation}
    \frac{d}{d\epsilon}\Big|_{\epsilon=0}\Gamma^A_{AB}(\epsilon)=\frac{3}{2}\frac{\partial \beta}{\partial X^B}.
\end{equation}
%-----------------------------
With these results, the linearized balance of linear momentum
(\ref{perturbed-linear-momentum1}) becomes
%-----------------------------
\begin{equation}
    \left(\mathbbm{A}^{aA}{}_b{}^BU^b{}_{,B}\right)_{,A}+\left(\mathbbm{B}^{aACD}G_{CD}~ \beta\right)_{,A}+\frac{3}{2}\frac{\partial \beta}{\partial X^B}P^{aB}=0.
\end{equation}
%-----------------------------
Assuming that $\boldsymbol{\mathbbm{A}}$ and
$\boldsymbol{\mathbbm{B}}$ are independent of $\mathbf{X}$, the
linearized equilibrium equations are simplified to read
%-----------------------------
\begin{equation}
    \mathbbm{A}^{aA}{}_b{}^B \frac{\partial^2 U^b}{\partial X^A\partial X^B}
    +\mathbbm{B}^{aACD}G_{CD} \frac{\partial \beta}{\partial X^A}+\frac{3}{2}\frac{\partial \beta}{\partial X^B}P^{aB}=0.
\end{equation}
%-----------------------------
If the initial configuration is stress-free, we have
%-----------------------------
\begin{equation}\label{linearized-momentum}
    \mathbbm{A}^{aA}{}_b{}^B \frac{\partial^2 U^b}{\partial X^A\partial X^B}
    =-\mathbbm{B}^{aACD}G_{CD} \frac{\partial \beta}{\partial X^A}.
\end{equation}
%-----------------------------
Let us next show that these results agree with those of classical
thermoelasticity. We first consider a special class of isotropic materials.

\paragraph{Saint-Venant-Kirchhoff materials.}
Saint-Venant-Kirchhoff materials have a constitutive relation that
is analogous to the linear isortropic materials, namely, the
second Piola-Kirchhoff stress $\mathbf{S}$ is given in terms of
the Lagrangian strain
$\mathbf{E}=\frac{1}{2}(\mathbf{C}-\mathbf{G})$ as
\citep{MaHu1983}
%-----------------------------
\begin{equation}
    \mathbf{S}=\lambda (\operatorname{tr}\mathbf{E})\mathbf{G}^{-1}+2\mu \mathbf{E},
\end{equation}
%-----------------------------
or in components
%-----------------------------
\begin{equation}\label{Saint-Venant-Kirchhoff}
    S^{CD}=\lambda E_{AB}G^{AB}G^{CD}+2\mu E^{CD}
    =\frac{\lambda}{2}(C_{AB}G^{AB}-3)G^{CD}+\mu(C_{AB}G^{AC}G^{BD}-G^{CD}),
\end{equation}
%-----------------------------
where $\lambda=\lambda(\mathbf{X})$ and $\mu=\mu(\mathbf{X})$ are
two scalars characterizing the material properties. This means
that $\mathbf{S}$ is a linear function of $\mathbf{E}$. We next
show that for this class of materials linearization of our
geometric theory leads to linear governing equations that are
identical to those of the classical linear theory of thermal
stresses for linear, isotropic materials. We will show later in
this section that this is true for any elastic material.

We can obtain the tensor $\mathbbm{B}^{aCAB}$ from $\mathbf{S}$ as
follows
%-----------------------------
\begin{equation}
    \mathbbm{B}^{aCAB}=\frac{\partial}{\partial G_{AB}}\left(g^{ab}\frac{\partial \psi}{\partial F^b{}_C}\right)
    =\frac{\partial P^{aC}}{\partial G_{AB}}=F^a{}_D\frac{\partial S^{CD}}{\partial G_{AB}}.
\end{equation}
%-----------------------------
Using
%-----------------------------
\begin{equation}
    \frac{\partial G^{AB}}{\partial G_{MN}}=-G^{AM}G^{BN},
\end{equation}
%-----------------------------
we obtain
%-----------------------------
\begin{equation}
    \mathbbm{B}^{aACD}G_{CD}=-2C_{MN}F^a{}_B\left(\lambda G^{AB}G^{MN}+2\mu G^{AM}G^{BN}\right)+(3\lambda+2\mu)F^a{}_BG^{AB}.
\end{equation}
%-----------------------------
The initial metric is Euclidean; in Cartesian coordinates,
$G_{AB}=\delta_{AB}$. Since the ambient space is also Euclidean,
we can choose a Cartesian coordinate system whose axes coincide
with the initial location of the material points along the
material Cartesian axis. This will give, $F^a{}_A=\delta^a_A$,
where $a$ and $A$ both range over $1, 2, 3$. This gives
%-----------------------------
\begin{equation}
    \mathbbm{B}^{aACD}G_{CD}=-\frac{3\lambda+2\mu}{2}~\delta^{aA}.
\end{equation}
%-----------------------------
Similarly, for an initially stress-free material manifold, we
obtain
%-----------------------------
\begin{equation}
    \mathbbm{A}^{aA}{}_b{}^B=F^a{}_MF^c{}_Ng_{bc}\left[\lambda G^{AM}G^{BN}+\mu(G^{AB}G^{MN}+G^{AN}G^{BM})\right].
\end{equation}
%-----------------------------
For the case of an initially Euclidean material manifold with
Cartesian coordinates we have
%-----------------------------
\begin{equation}
    \mathbbm{A}^{aA}{}_b{}^B \frac{\partial^2 U^b}{\partial X^A\partial X^B}
    =(\lambda+\mu)U_{b,ab}+\mu U_{a,bb}.
\end{equation}
%-----------------------------
Therefore, Eq. (\ref{linearized-momentum}) reads
%-----------------------------
\begin{equation}
    (\lambda+\mu)U_{b,ab}+\mu U_{a,bb}=\frac{3\lambda+2\mu}{2}\frac{\partial \beta}{\partial x_a}.
\end{equation}
%-----------------------------
Recalling $\beta=2\alpha \frac{d T}{d\epsilon}\big|_{\epsilon=0}$,
and assuming for simplicity that $\alpha$ is constant, we have
%-----------------------------
\begin{equation}
    \frac{\partial \beta}{\partial
    x_a}=2\alpha\frac{\partial}{\partial x_a}\frac{d
    T}{d\epsilon}\Big|_{\epsilon=0}.
\end{equation}
%-----------------------------
Hence
%-----------------------------
\begin{equation}\label{nonlinear-linear-momentum}
    (\lambda+\mu)U_{b,ab}+\mu
    U_{a,bb}=(3\lambda+2\mu)\alpha\frac{\partial}{\partial
    x_a}\frac{dT}{d\epsilon}\Big|_{\epsilon=0}.
\end{equation}
%-----------------------------

In the classical theory of thermal stresses, stress-strain
relations in the presence of temperatures changes can be written
as
%-----------------------------
\begin{equation}\label{classical-stress}
    \sigma_{ij}=\textsf{C}_{ijkl}(\epsilon_{kl}-\alpha \delta_{kl}\Delta T),
\end{equation}
%-----------------------------
where $\textsf{C}_{ijkl}$ is the elasticity tensor. In this sense
thermal strains are understood as ``eigen strains". Equilibrium
equations in the absence of body forces read
%-----------------------------
\begin{equation}
    \textsf{C}_{ijkl}\epsilon_{kl,j}=\textsf{C}_{ijkk}\alpha \frac{\partial\Delta T}{\partial
    x_j},
\end{equation}
%-----------------------------
where once again we have assumed that the elasticity tensor and
the coefficient of thermal expansion are constants. When the
material is isotropic, $\textsf{C}_{ijkl}=\mu(
\delta_{ik}\delta_{jl}+\delta_{il}\delta_{jk})+\lambda\delta_{ij}\delta_{kl}$
and hence
%-----------------------------
\begin{equation}
    \textsf{C}_{ijkl}\epsilon_{kl,j}=\mu
    u_{i,jj}+(\lambda+\mu)u_{j,ji},
\end{equation}
%-----------------------------
and
%-----------------------------
\begin{equation}
    \textsf{C}_{ijkk}\alpha \frac{\partial\Delta T}{\partial x_j}=(2\mu+3\lambda)\alpha\frac{\partial\Delta T}{\partial x_i}.
\end{equation}
%-----------------------------
Thus, equilibrium equations read
%-----------------------------
\begin{equation}\label{linear-linear-momentum}
    \mu u_{i,jj}+(\lambda+\mu)u_{j,ji}=(2\mu+3\lambda)\alpha\frac{\partial\Delta T}{\partial x_i}.
\end{equation}
%-----------------------------
Recalling that $\frac{d T}{d\epsilon}\big|_{\epsilon=0}$ is the
linearized version of temperature change, i.e., $\delta T \approx
\epsilon\frac{dT}{d\epsilon}\big|_{\epsilon=0}$ and that
$\mathbf{U}$ is the linearized version of displacement, it is seen
that the linearization of the geometric theory for
Saint-Venant-Kirchhoff materials results in governing equations
that are identical to those of the classical isotropic linear
theory.

Let us now see if this holds for more general constitutive
equations of the geometric theory. For a stress-free (Euclidean)
initial configuration, the linearized balance of linear momentum
is given by
%-----------------------------
\begin{equation}\label{classic-relation}
    \left(\mathbbm{A}^{aA}{}_b{}^B~U^b_{,B}+\mathbbm{B}^{aACD}\delta_{CD}~\beta \right)_{,A}=0.
\end{equation}
%-----------------------------
Assuming  $G_{AB}=\delta_{AB}$ and $F^a{}_A=\delta^a_A$ as above,
an identity proven in \citep{MaHu1983} becomes
%-----------------------------
\begin{equation}
    \mathbbm{A}^{aA}{}_b{}^B=2\mathbbm{C}^{CADB}F^c{}_DF^a{}_C g_{cb}=2\mathbbm{C}^{CADB}\delta^c_D\delta^a_C\delta_{bc}=2\mathbbm{C}^{aAbB}.
\end{equation}
%-----------------------------
Noting that $\beta=2\alpha \frac{d
T}{d\epsilon}\big|_{\epsilon=0}$, (\ref{classic-relation}) becomes
%-----------------------------
\begin{equation}
    \left(\mathbbm{C}^{aAbB}~U^b_{,B}+\mathbbm{B}^{aACC}
    \alpha~\frac{dT}{d\epsilon}\Big|_{\epsilon=0} \right)_{,A}=0.
\end{equation}
%-----------------------------
Identifying superscripts and subscripts, identifying the spatial
and material indices (by aligning the material and spatial
Cartesian coordinates as above), and using symmetries of
$\mathbbm{C}$, we can write
%-----------------------------
\begin{equation}
    \left(\mathbbm{C}_{ljki} \epsilon_{ki}+\mathbbm{B}_{ljkk} \alpha~
    \frac{dT}{d\epsilon}\Big|_{\epsilon=0} \right)_{,j}=0.
\end{equation}
%-----------------------------
This is identical to (\ref{classical-stress}) if
%-----------------------------
\begin{equation}\label{B-C-condition}
    \mathbbm{B}_{ljkk}=-\mathbbm{C}_{ljki}\delta_{ki}=-\mathbbm{C}_{ljkk}.
\end{equation}
%-----------------------------

Let us first show that this relation always holds for isotropic
materials. For isotropic materials it can be shown that
\citep{LuPapadopoulos2000,LuPapadopoulos2003,YaMaOr2006}
%-----------------------------
\begin{equation}\label{infinitesimal-covariance}
    \frac{\partial\Psi}{\partial \mathbf{C}}\cdot\mathbf{C}+\frac{\partial\Psi}{\partial \mathbf{G}}\cdot\mathbf{G}=\mathbf{0}.
\end{equation}
%-----------------------------
Or in components
%-----------------------------
\begin{equation}
    \frac{\partial\Psi}{\partial C_{AC}}C_{CB}+\frac{\partial\Psi}{\partial G_{AC}}G_{CB}=0.
\end{equation}
%-----------------------------
Note that
%-----------------------------
\begin{equation}
    \boldsymbol{\mathbbm{B}}=\frac{\partial \mathbf{P}}{\partial \mathbf{G}}
    =\mathbf{F}\frac{\partial \mathbf{S}}{\partial \mathbf{G}}
    =2\mathbf{F}\frac{\partial^2 \Psi}{\partial \mathbf{C} \partial \mathbf{G}}.
\end{equation}
%-----------------------------
Differentiating (\ref{infinitesimal-covariance}) with respect to
$\mathbf{C}$ and noting that the initial configuration is stress
free we obtain
%-----------------------------
\begin{equation}
    \frac{\partial^2 \Psi}{\partial \mathbf{C} \partial \mathbf{C}}\cdot\mathbf{C}
    +\frac{\partial^2 \Psi}{\partial \mathbf{C} \partial \mathbf{G}}\cdot\mathbf{G}
    =\mathbf{0}.
\end{equation}
%-----------------------------
Or
%-----------------------------
\begin{equation}\label{covariance-simplified}
    \frac{1}{2}\boldsymbol{\mathbbm{C}}\cdot\mathbf{C}+\frac{1}{2}\mathbf{F}^{-1}\boldsymbol{\mathbbm{B}}\cdot\mathbf{G}
    =\mathbf{0}.
\end{equation}
%-----------------------------
Noting that $F^a{}_A=\delta^a_A$ and $C_{AB}=G_{AB}$,
(\ref{covariance-simplified}) is identical to
(\ref{B-C-condition}).

We now show that Eq.(\ref{B-C-condition}) holds even for
anisotropic elastic solids. For showing this we use the fact that
if the material is homogeneous (i.e., if the thermal expanstion
properties do not depend on position) a uniform temperature change
$\Delta T$ does not lead to any thermal stresses. Starting from a
stress-free Euclidean configuration, for a uniform temperature
change, one has
%-----------------------------
\begin{equation}
    \delta F^a{}_A=U^a{}_{,A}=\alpha \Delta T \delta^a{}_A~~~~~\textrm{and}~~~~~\delta \mathbf{S}=\mathbf{0}.
\end{equation}
%-----------------------------
We also know $\mathbf{S}=\mathbf{S}(\mathbf{C},\mathbf{G})$, thus
%-----------------------------
\begin{equation}\label{delta-S}
    \delta \mathbf{S}=\frac{\partial \mathbf{S}}{\partial\mathbf{C}}\cdot\delta\mathbf{C}+\frac{\partial \mathbf{S}}{\partial\mathbf{G}}\cdot\delta\mathbf{G}
    =\boldsymbol{\mathbbm{C}}\cdot\delta\mathbf{C}+2\alpha\Delta T ~\mathbf{F}^{-1}\boldsymbol{\mathbbm{B}}\cdot\mathbf{G}=\mathbf{0}.
\end{equation}
%-----------------------------
But note that \citep{YavariOzakin2008}
%-----------------------------
\begin{equation}\label{delta-C}
    \delta C_{AB}=g_{ab}F^a{}_A U^b{}_{|B}+g_{ab}F^b{}_B U^a{}_{|A}=\delta_{ab}\delta^a_A(\alpha \Delta T \delta^b_B)+\delta_{ab}\delta^b_B(\alpha \Delta T \delta^a_A)
    =2\alpha \Delta T \delta_{AB}.
\end{equation}
%-----------------------------
Substituting (\ref{delta-C}) into (\ref{delta-S}), one obtains
(\ref{B-C-condition})! In summary, we have proved the following
proposition.

\paragraph{Proposition.} Linearization of the present geometric theory yields governing
equations that are identical to those of classical linear
elasticity.

\section{Conclusions}

In this paper, we presented a geometric theory of thermal stresses
in which the material manifold is temperature dependent. Given a
temperature distribution, the material metric is a Riemannian
metric that is obtained by a (non-uniform) rescaling of a
reference metric. In particular, starting from a Euclidean
stress-free reference manifold, a non-uniform temperature
distribution leads to a non-Euclidean material manifold. We
studied the stress-free temperature distributions by looking at
conditions that guarantee flatness of a Riemannian metric. We
recovered some known facts from the linear theory of thermal
stresses and obtained some new results for finite deformations. We
showed that, in addition to uniform temperature distributions,
there are other zero-stress temperature distributions. We obtained
all such temperature distributions. We also studied the inverse
problem, i.e., given a temperature distribution, what
inhomogeneous coefficients of thermal expansion give zero
stresses. In the present theory, there is no need to introduce an
``intermediate" configuration. We made an explicit connection
between our geometric theory and the previous works on
multiplicative decomposition of deformation gradient in the
presence of temperature changes. Given a temperature distribution,
we obtained the temperature-dependent governing equations. In
order to demonstrate the power of the geometric theory, we solved
the example of an axisymmetric temperature distribution and
obtained some exact results. We showed that linearization of the
present geometric theory about a stress-free configuration results
in governing equations that are identical to those of the
classical linear thermoelasticity.

Geometric formulation of the coupled problem of elastic
deformations with heat conduction will be studied in a future
communication. The ideas presented in this paper can also be used
in modeling bodies with growing mass. Growth and remodeling in
biological systems is an important phenomenon and a geometric
study will shed light on the coupling between growth/remodeling
and elastic deformations.

%%%%%%%%%%%%%%%%%%%%%%%%%%%%%%%%%%%%%%%%%%%%%%

\appendix
\section{Differential Geometry and Classical Geometric Elasticity}

In this section, in order to make the paper self-contained, we
review some notation from geometric elasticity. For more details
refer to \citep{MaHu1983, AbMaRa1988, Mars03}. By \emph{classical}
geometric elasticity we mean elasticity of bodies with stationary
defects (if any) and a fixed material manifold. We extended this
theory for thermal deformations in \S3.

For a smooth $n$-manifold $M$, the tangent space to $M$ at a point
$p \in M$  is denoted $T_pM$ and the whole tangent bundle is
denoted $TM$. We denote by $\mathcal{B}$ a reference manifold for
our body and by $\mathcal{S}$ the space in which the body moves.
We assume that $\mathcal{B}$ and $\mathcal{S}$ are Riemannian
manifolds with metrics $\mathbf{G}$ and $\mathbf{g}$,
respectively. Local coordinates on $\mathcal{B}$ are denoted by
$\{X ^A\}$ and those on $\mathcal{S}$ by $\{x ^a\}$.

A {\bfi deformation} of the body is a $C ^1$  embedding
$\varphi:\mathcal{B}\rightarrow \mathcal{S}$. The tangent map of
$\varphi $ is denoted $\mathbf{F} = T \varphi: T \mathcal{B}
\rightarrow T \mathcal{S}$, which is often called the deformation
gradient. In local charts on $\mathcal{B} $ and $\mathcal{S}$, the
tangent map of $ \varphi$ is given by the Jacobian matrix of
partial derivatives of the components of $\varphi$, as
%-----------------------------
\begin{equation}
    \mathbf{F} = T\varphi:T\mathcal{B}\rightarrow
   T\mathcal{S},~~~T\varphi(\mathbf{X},\mathbf{Y})=(\varphi(\mathbf{X}),\mathbf{D}\varphi(\mathbf{X}) \cdot
    \mathbf{Y}).
\end{equation}
%-----------------------------

If $\mathbf{Y}$ is a vector field on $\mathcal{B}$, then
$\varphi_*\mathbf{Y} = T\varphi \cdot \mathbf{Y} \circ
\varphi^{-1}$, or using the $\mathbf{F}$ notation,
$\varphi_*\mathbf{Y} = \mathbf{F} \cdot \mathbf{Y} \circ
\varphi^{-1}$ is a vector field on $\varphi(\mathcal{B})$ called
the {\bfi push-forward} of $\mathbf{Y}$ by $\varphi$. Similarly,
if $\mathbf{y}$ is a vector field on $\varphi(\mathcal{B}) \subset
\mathcal{S}$, then $\varphi^*\mathbf{y}=T(\varphi^{-1}) \cdot
\mathbf{y} \circ \varphi$ is a vector field on $\mathcal{B}$ and
is called the pull-back of $\mathbf{y}$ by $\varphi$.

The cotangent bundle of a manifold $M$ is denoted $T ^{\ast} M $
and the fiber at a point $p \in M $ (the vector space of one-forms
at $p$) is denoted by $T_p^*M$. If $\mathbf{\beta}$ is a one-form
on $\mathcal{S}$, i.e., a section of the cotangent bundle $ T
^{\ast} \mathcal{S}$, then the one-form on $\mathcal{B}$ defined
as
%-----------------------------
\begin{equation}
    (\varphi^{*}\mathbf{\beta})_{\mathbf{X}} \cdot
    \mathbf{V}_{\mathbf{X}}=\mathbf{\beta}_{\varphi(\mathbf{X})} \cdot (T\varphi \cdot
    \mathbf{V}_{\mathbf{X}}) =\mathbf{\beta}_{\varphi(\mathbf{X})} \cdot (\mathbf{F} \cdot
    \mathbf{V}_{\mathbf{X}})
    \end{equation}
%-----------------------------
for $\mathbf{X} \in \mathcal{B} $ and $\mathbf{V}_{\mathbf{X}} \in
T_{\mathbf{X}}\mathcal{B}$, is called the {\bfi pull-back} of
$\mathbf{\beta}$ by $\varphi$. Similarly, the {\bfi push-forward}
of a one-form $\mathbf{\alpha}$ on $\mathcal{B}$ is the one form
on $\varphi ( \mathcal{B} ) $ defined by
$\varphi_{*}\mathbf{\alpha}=(\varphi^{-1})^*\mathbf{\alpha}$.

We can associate a vector field $\beta ^{\sharp} $ to a one-form
$\beta$ on a Riemannian manifold $M$ through the equation
%-----------------------------
\begin{equation}
     \left\langle \beta_{\mathbf{x}}, \mathbf{v}_{\mathbf{x}} \right\rangle =
\left\langle \!\! \left\langle \beta^{\sharp}_{\mathbf{x}},
\mathbf{v}_{\mathbf{x}} \right\rangle \!\!
\right\rangle_{\mathbf{x}},
\end{equation}
%-----------------------------
where $\left\langle \, , \right\rangle$ denotes the natural
pairing between the one form $\beta _{\mathbf{x}} \in T^*
_{\mathbf{x}} M $ and the vector $\mathbf{v}_{\mathbf{x}} \in T
_{\mathbf{x}} M$ and where $\left\langle \!\! \left\langle
\beta^{\sharp}_{\mathbf{x}}, \mathbf{v}_{\mathbf{x}} \right\rangle
\!\! \right\rangle_{\mathbf{x}} $ denotes the inner product
between $\beta ^{\sharp} _{\mathbf{x}} \in T _{\mathbf{x}} M$ and
$\mathbf{v}_{\mathbf{x}} \in T _{\mathbf{x}} M$ induced by the
metric $\mathbf{g}$. In coordinates, the components of $\beta
^{\sharp}$ are given by $\beta ^a = g ^{ab} \beta _b$.

A type $\begin{pmatrix} m \\
n \end{pmatrix}$-tensor at $\mathbf{X} \in \mathcal{B}$ is a
multilinear map
%-----------------------------
\begin{equation}
    \mathbf{T}:\underbrace{T_{\mathbf{X}}^{*}\mathcal{B}\times ... \times T_{\mathbf{X}}^{*}\mathcal{B}}_{m~\textrm{copies}}\times
    \underbrace{T_{\mathbf{X}}\mathcal{B}\times ... \times T_{\mathbf{X}}\mathcal{B}}_{n~\textrm{copies}} \rightarrow
    \mathbb{R}.
\end{equation}
%-----------------------------
$\mathbf{T}$ is said to be contravariant of order $m$ and
covariant of order $n$. In a local coordinate chart
%-----------------------------
\begin{equation}
    \mathbf{T}(\mathbf{\alpha}^1,...,\mathbf{\alpha}^m,\mathbf{V}_1,...,\mathbf{V}_n)=
    T^{~i_1...i_m}{}_{j_1...j_n}\alpha^1_{i_1}...\alpha^m_{i_m}V_1^{j_1}...V_n^{j_n},
\end{equation}
%-----------------------------
where $\mathbf{\alpha}^k \in T_{\mathbf{X}}^* \mathcal{B}$ and
$\mathbf{V}^k \in T_{\mathbf{X}} \mathcal{B}$.

A {\bfi two-point tensor} $\mathbf{T}$ of type $\begin{pmatrix}  m & r \\
  n & s \end{pmatrix}$ at $\mathbf{X}\in \mathcal{B}$ over a map
$\varphi:\mathcal{B}\rightarrow \mathcal{S}$ is a multilinear map
%-----------------------------
\begin{equation}
    T:\underbrace{T_{\mathbf{X}}^{*}\mathcal{B}\times ... \times T_{\mathbf{X}}^{*}\mathcal{B}}_{m~\textrm{copies}}\times
    \underbrace{T_{\mathbf{X}}\mathcal{B}\times ... \times T_{\mathbf{X}}\mathcal{B}}_{n~\textrm{copies}}
    \times\underbrace{T_{\mathbf{x}}^{*}\mathcal{S}\times ... \times T_{\mathbf{x}}^{*}\mathcal{S}}_{r~\textrm{copies}}\times
    \underbrace{T_{\mathbf{x}}\mathcal{S}\times ... \times T_{\mathbf{x}}\mathcal{S}}_{s~\textrm{copies}}    \rightarrow
    \mathbb{R},
\end{equation}
%-----------------------------
where $\mathbf{x}=\varphi(\mathbf{X})$.

Let $\mathbf{y}$ be a vector field on $\mathcal{S}$ and
$\varphi:\mathcal{B} \rightarrow \mathcal{S}$ a regular and
orientation preserving $C^1$ map. The {\bfi Piola transform} of
$\mathbf{y}$ is defined as
%-----------------------------
\begin{equation}
    \mathbf{Y}=J \varphi^* \mathbf{y},
\end{equation}
%-----------------------------
where $J$ is the Jacobian of $\varphi$. If $\mathbf{Y}$ is the
Piola transform of $\mathbf{y}$, then the {\bfi Piola identity}
holds:
%-----------------------------
\begin{equation}
    \operatorname{Div}\mathbf{Y} = J(\operatorname{div}\mathbf{y}) \circ
    \varphi.
\end{equation}
%-----------------------------

A $p$-form on a manifold $M$ is a skew-symmetric $\begin{pmatrix}  0 \\
  p \end{pmatrix}$-tensor. The space of $p$-forms on $M$ is denoted
by $\Omega^p(M)$. If $\varphi:M\rightarrow N$ is a regular and
orientation preserving $C^1$ map and $\mathbf{\alpha} \in
\Omega^p(\varphi(M))$, then
%-----------------------------
\begin{equation}
    \int_{\varphi(M)}\mathbf{\alpha}=\int_{M}\varphi^*\mathbf{\alpha}.
\end{equation}
%-----------------------------

Let $\pi:E\rightarrow \mathcal{S}$ be a vector bundle over a
manifold $\mathcal{S}$ and let $\mathcal{E}(\mathcal{S})$ be the
space of smooth sections of $E$ and $\mathcal{X}(\mathcal{S})$ the
space of vector fields on $\mathcal{S}$. A
\emph{\textbf{connection}} on $E$ is a map
$\nabla:\mathcal{X}(\mathcal{S})\times\mathcal{E}(\mathcal{S})\rightarrow\mathcal{E}(\mathcal{S})$
such that $\forall~f,f_1,f_2\in
C^{\infty}(\mathcal{S}),~\forall~a_1,a_2\in\mathbb{R}$
%-----------------------------
\begin{eqnarray}
  & i)& \nabla_{f_1\mathbf{X}_1+f_2\mathbf{X}_2}\mathbf{Y}=f_1\nabla_{\mathbf{X}_1}\mathbf{Y}
        +f_2\nabla_{\mathbf{X}_2}\mathbf{Y}, \\
  & ii)& \nabla_{\mathbf{X}}(a_1\mathbf{Y}_1+a_2\mathbf{Y}_2)=a_1\nabla_{\mathbf{X}}(\mathbf{Y}_1)
        +a_2\nabla_{\mathbf{X}}(\mathbf{Y}_2), \\
  & iii)&
  \nabla_{\mathbf{X}}(f\mathbf{Y})=f\nabla_{\mathbf{X}}\mathbf{Y}+(\mathbf{X}f)\mathbf{Y}.
\end{eqnarray}
%-----------------------------
A \emph{\textbf{linear connection}} on $\mathcal{S}$ is a
connection on $T\mathcal{S}$, i.e.,
$\nabla:\mathcal{X}(\mathcal{S})\times\mathcal{X}(\mathcal{S})\rightarrow\mathcal{X}(\mathcal{S})$.
In a local chart
%-----------------------------
\begin{equation}
    \nabla_{\partial_i}\partial_j=\gamma_{ij}^k
    \partial_k,
\end{equation}
%-----------------------------
where $\gamma_{ij}^k$ are Christoffel symbols of the connection
and $\partial_i=\frac{\partial}{\partial x^i}$. A linear
connection is said to be compatible with the metric of the
manifold if
%-----------------------------
\begin{equation}
    \nabla_{\mathbf{X}}\left\langle\!\left\langle \mathbf{Y},\mathbf{Z}\right\rangle\!\right\rangle
    =\left\langle\!\left\langle \nabla_{\mathbf{X}}\mathbf{Y},\mathbf{Z}\right\rangle\!\right\rangle
    +\left\langle\!\left\langle \mathbf{Y},\nabla_{\mathbf{X}}\mathbf{Z}\right\rangle\!\right\rangle.
\end{equation}
%-----------------------------
It can be shown that $\nabla$ is compatible with
$\mathbf{g}$ if and only if
$\nabla\mathbf{g}=\mathbf{0}$.
\emph{\textbf{Torsion}} of a connection is defined as
%-----------------------------
\begin{equation}
    \boldsymbol{\mathcal{T}}(\mathbf{X},\mathbf{Y})=\nabla_{\mathbf{X}}\mathbf{Y}-\nabla_{\mathbf{Y}}\mathbf{X}
    -[\mathbf{X},\mathbf{Y}],
\end{equation}
%-----------------------------
where
%-----------------------------
\begin{equation}
    [\mathbf{X},\mathbf{Y}](F)=\mathbf{X}(\mathbf{Y}(F))-\mathbf{Y}(\mathbf{X}(F))~~~~~\forall~F\in C^{\infty}(\mathcal{S}),
\end{equation}
%-----------------------------
is the \emph{\textbf{commutator}} of $\textbf{X}$ and
$\textbf{Y}$. $\nabla$ is symmetric if it is
torsion-free, i.e.
%-----------------------------
\begin{equation}
    \nabla_{\mathbf{X}}\mathbf{Y}-\nabla_{\mathbf{Y}}\mathbf{X}
    =[\mathbf{X},\mathbf{Y}].
\end{equation}
%-----------------------------
It can be shown that on any Riemannian manifold
$(\mathcal{S},\mathbf{g})$ there is a unique linear connection
$\nabla$ that is compatible with $\mathbf{g}$ and is
torsion-free with the following Christoffel symbols
%-----------------------------
\begin{equation}
    \gamma_{ij}^k=\frac{1}{2}g^{kl}\left(\frac{\partial g_{jl}}{\partial x^i}+\frac{\partial g_{il}}{\partial x^j}
    -\frac{\partial g_{ij}}{\partial x^l}\right).
\end{equation}
%-----------------------------
This is the \textbf{\emph{Fundamental Lemma of Riemannian
Geometry}} \citep{Lee1997} and this connection is called the
\emph{\textbf{Levi-Civita connection}}.

\emph{\textbf{Curvature tensor}} $\boldsymbol{\mathcal{R}}$ of a
Riemannian
manifold $(\mathcal{S},\mathbf{g})$ is a $\begin{pmatrix}  1 \\
  3 \end{pmatrix}$-tensor $\boldsymbol{\mathcal{R}}:T^*_{\mathbf{x}}\mathcal{S}\times
T_{\mathbf{x}}\mathcal{S}\times T_{\mathbf{x}}\mathcal{S}\times
T_{\mathbf{x}}\mathcal{S}\rightarrow\mathbb{R}$ defined as
%-----------------------------
\begin{equation}
    \boldsymbol{\mathcal{R}}(\alpha,\mathbf{w}_1,\mathbf{w}_2,\mathbf{w}_3)=\alpha\left(\nabla_{\mathbf{w}_1}\nabla_{\mathbf{w}_2}\mathbf{w}_3
    -\nabla_{\mathbf{w}_2}\nabla_{\mathbf{w}_1}\mathbf{w}_3
    -\nabla_{[\mathbf{w}_1,\mathbf{w}_2]}\mathbf{w}_3\right)
\end{equation}
%-----------------------------
for $\alpha\in
T^*_{\mathbf{x}}S,~\mathbf{w}_1,\mathbf{w}_2,\mathbf{w}_3\in
T_{\mathbf{x}}S$. In a coordinate chart $\{x^a\}$
%-----------------------------
\begin{equation}
    \mathcal{R}^a{}_{bcd}=\frac{\partial \gamma^a_{bd}}{\partial x^c}-\frac{\partial \gamma^a_{bc}}{\partial x^d}
    +\gamma^a_{ce}\gamma^e_{bd}-\gamma^a_{de}\gamma^e_{bc}.
\end{equation}
%-----------------------------
Note that for an arbitrary vector field $\mathbf{w}$
%-----------------------------
\begin{equation}
    w^a{}_{|bc}-w^a{}_{|cb}=\mathcal{R}^a{}_{bcd}w^d+\mathcal{T}^d{}_{cb}w^a{}_{|d}.
\end{equation}
%-----------------------------
An $n$-dimensional Riemannian manifold is flat if it is isometric
to Euclidean space. A Riemannian manifold is flat if and only if
its curvature tensor vanishes. A Riemannian manifold
$(\mathcal{B},\mathbf{G})$ is conformally flat if there exists a
smooth map $f:\mathcal{B}\rightarrow\mathbb{R}$ such that
$\mathbf{G}=f\boldsymbol{\delta}$, where $\boldsymbol{\delta}$ is
the Euclidean metric. In \emph{isothermal coordinates} the
Riemannian metric has the following local form
%-----------------------------
\begin{equation}
    \mathbf{G}=f(\mathbf{X})\left(dX_1^2+...+dX_n^2\right).
\end{equation}
%-----------------------------
It is know that \citep{Berger2003} any two-dimensional Riemannian
manifold is conformally flat and the map $f$ is unique. A
corollary of this theorem in our theory of thermal stresses is the
following. Given any smooth curved 2D solid that is stress free,
there exists a unique change of temperature distribution such that
in the new temperature distribution the 2D solid is flat and still
stress free. Equivalently, starting from a stress free flat sheet,
it is always possible to deform it to any smooth curved shape by
changing temperature without imposing any residual stresses.

For a Riemannian manifold $(\mathcal{B},\mathbf{G})$ the
Weyl-Schouten tensor is defined as \citep{Nakahara2003}
%-----------------------------
\begin{equation}
    \textsf{C}_{IJK}=\nabla_K\mathcal{R}_{IJ}-\nabla_J\mathcal{R}_{IK}
    -\frac{1}{4}\left(G_{IJ}\frac{\partial\mathcal{R}}{\partial X^K}-G_{IK}\frac{\partial\mathcal{R}}{\partial
    X^J}\right),
\end{equation}
%-----------------------------
where $\boldsymbol{\mathcal{R}}$ and $\mathcal{R}$ are the Ricci
tensor and the scalar curvature, respectively. A necessary and
sufficient condition for a Riemannian manifold
$(\mathcal{B},\mathbf{G})$ to be conformally flat is
$\boldsymbol{\textsf{C}}=\boldsymbol{\textsf{0}}$ when $\dim
\mathcal{B}=3$.

\subsection{Absolute Parallelizable (AP) Geometry}

In many physical problems in which deformation is coupled with
other phenomena, e.g. plasticity, growth/remodeling, thermal
expansion/contraction, etc. all one can hope to do is to locally
decouple the elastic deformations from the inelastic deformations.
This has led to many works that start from a decomposition of
deformation gradient $\mathbf{F}=\mathbf{F}_e\mathbf{F}_i$, where
$\mathbf{F}_e$ is the elastic deformation gradient and
$\mathbf{F}_i$ is the remaining local deformation or inelastic
deformation gradient.

Given an inelastic deformation gradient, here a thermal
deformation gradient, a vector in the tangent space of
$\mathbf{X}\in\mathcal{B}$, i.e. $\mathbf{W}\in
T_{\mathbf{X}}\mathcal{B}$ is mapped to another vector
$\hat{\mathbf{W}}=\mathbf{F}_T\mathbf{W}$. Traditionally these
vectors are assumed to lie in the tangent bundle of an
``intermediate configuration." In the literature, intermediate
configuration is not clearly defined and at first glance it seems
to be more or less mysterious as was explained in \S 2.3. These
are closely related to parallelizable manifolds (or absolutely
parallelizable (AP) manifolds) going back to the works by
\citet{Eisenhart1926,Eisenhart1927}. See also
\citet{YoussefSid-Ahmed2007} and \citet{Wanas2008}. In an
$n$-dimensional AP-manifold $M$, one starts with a field of $n$
linearly independent vectors $\left\{\mathbf{E}_{(A)} \right\}$
that span the tangent vector at each point. Let us denote the
components of $\mathbf{E}_{(A)}$ by $\mathbf{E}_{(A)}^I$. The dual
vectors, i.e. the corresponding basis vectors for the cotangent
space are denoted by $\left\{\mathbf{E}^{(A)} \right\}$ with
components $\left\{\mathbf{E}^{(A)}_I \right\}$. Note that
%-----------------------------
\begin{equation}
    \mathbf{E}^{(A)}_I\mathbf{E}_{(B)}^I=\delta^A_B~~~~~\textrm{and}~~~~~\mathbf{E}^{(A)}_I\mathbf{E}_{(A)}^J=\delta^I_J.
\end{equation}
%-----------------------------
One is then interested in equipping $M$ with a connection
$\Gamma^I_{JK}$ such that the basis vectors
$\left\{\mathbf{E}_{(A)} \right\}$ are covariantly constant,
i.e.\footnote{Equivalently, the tangent bundle is a trivial
bundle, so that the associated principal bundle of linear frames
has a section on $M$.}
%-----------------------------
\begin{equation}\label{parallel}
    \mathbf{E}_{(A)}^I{}_{|J}=0.
\end{equation}
%-----------------------------
Note that
%-----------------------------
\begin{equation}
    \mathbf{E}_{(A)}^I{}_{|JK}-\mathbf{E}_{(A)}^I{}_{|KJ}=\mathcal{R}^I{}_{LJK}\mathbf{E}_{(A)}^L+\mathcal{T}^L{}_{KJ}\mathbf{E}_{(A)}^I{}_{|L}.
\end{equation}
%-----------------------------
Thus (\ref{parallel}) implies that
%-----------------------------
\begin{equation}
    \mathcal{R}^I{}_{LJK}=0
\end{equation}
%-----------------------------
i.e., $M$ is flat with respect to the connection $\Gamma^I_{JK}$.
Note that
%-----------------------------
\begin{equation}
    \mathbf{E}_{(A)}^I{}_{|J}=\frac{\partial \mathbf{E}_{(A)}^I}{\partial
    X^J}+\Gamma^I_{JK}\mathbf{E}_{(A)}^K.
\end{equation}
%-----------------------------
Thus
%-----------------------------
\begin{equation}
    \mathbf{E}^{(A)}_L\frac{\partial \mathbf{E}_{(A)}^I}{\partial X^J}+\Gamma^I_{LK}=0.
\end{equation}
%-----------------------------
Hence
%-----------------------------
\begin{equation}
    \Gamma^I_{JK}=-\mathbf{E}^{(A)}_J\frac{\partial \mathbf{E}_{(A)}^I}{\partial X^K}=\mathbf{E}^{(A)}_I\frac{\partial \mathbf{E}^{(A)}_J}{\partial X^K}.
\end{equation}
%-----------------------------
This is similar in form to the connections used by many authors,
e.g. by \citet{BilbyBulloughSmith1955} and \citet{Kondo1955a} for
dislocations, by \citet{EpsteinElzanowski2007} for material
inhomogeneities, and by \citet{Stojanovic1964} for thermal
stresses.

Looking at local charts $\left\{X^A\right\}$ and
$\left\{U^I\right\}$ for the reference and intermediate
configurations, we have
%-----------------------------
\begin{equation}
    dU^I=\left(F_T\right)^I{}_A dX^A.
\end{equation}
%-----------------------------
$\left(F_T\right)^I{}_A$ can be identified with
$\mathbf{E}_{(A)}^I$, and hence
%-----------------------------
\begin{equation}
    \Gamma^I_{JK}=\left(F_T\right)^I{}_A\frac{\partial \left(F_T^{-1}\right)^A{}_J}{\partial
    X^K},
\end{equation}
%-----------------------------
is the connection used in \citep{Stojanovic1964,Stojanovic1969}.
Note that this connection is curvature free but has non-vanishing
torsion.

\subsection{Geometric Elasticity}

Let us next review a few of the basic notions of geometric
continuum mechanics. A {\bfi body} $\mathcal{B}$ is identified
with a Riemannian manifold $\mathcal{B}$ and a {\bfi
configuration} of $\mathcal{B}$ is a mapping $\varphi:\mathcal{B}
\rightarrow \mathcal{S}$, where $\mathcal{S}$ is another
Riemannian manifold. The set of all configurations of
$\mathcal{B}$ is denoted $\mathcal{C}$. A {\bfi motion} is a curve
$c:\mathbb{R}\rightarrow \mathcal{C}; t\mapsto \varphi_t $ in
$\mathcal{C}$. It is assumed that the body is stress free in the
material manifold.

For a fixed $t$, $\varphi_t(\mathbf{X})=\varphi(\mathbf{X},t)$ and
for a fixed $\mathbf{X}$,
$\varphi_{\mathbf{X}}(t)=\varphi(\mathbf{X},t)$, where
$\mathbf{X}$ is position of material points in the undeformed
configuration $\mathcal{B}$. The {\bfi material velocity} is the
map $\mathbf{V}_t:\mathcal{B}\rightarrow \mathbb{R}^3$ given by
%-----------------------------
\begin{equation}
    \mathbf{V}_t(\mathbf{X})=\mathbf{V}(\mathbf{X},t)=\frac{\partial \varphi(\mathbf{X},t)}{\partial
    t}=\frac{d}{dt}\varphi_{\mathbf{X}}(t).
\end{equation}
%-----------------------------
Similarly, the {\bfi material acceleration} is defined by
%-----------------------------
\begin{equation}
    \mathbf{A}_t(\mathbf{X})=\mathbf{A}(\mathbf{X},t)=\frac{\partial \mathbf{V}(\mathbf{X},t)}{\partial
    t}=\frac{d}{dt}\mathbf{V}_{\mathbf{X}}(t).
\end{equation}
%-----------------------------
In components
%-----------------------------
\begin{equation}
    A^a=\frac{\partial V^a}{\partial t}+\gamma^a_{bc}V^bV^c,
\end{equation}
%-----------------------------
where $\gamma^a_{bc}$ is the Christoffel symbol of the local
coordinate chart $\{x^a\}$. Note that $\mathbf{A}$ does not depend
on the connection coefficients of the material manifold.

Here it is assumed that $\varphi_t$ is invertible and regular. The
{\bfi spatial velocity} of a regular motion $\varphi_t$ is defined
as
%-----------------------------
\begin{equation}
    \mathbf{v}_t:\varphi_t(\mathcal{B})\rightarrow
    \mathbb{R}^3,~~~~\mathbf{v}_t=\mathbf{V}_t\circ
    \varphi_t^{-1},
\end{equation}
%-----------------------------
and the {\bfi spatial acceleration} $\mathbf{a}_t$ is defined as
%-----------------------------
\begin{equation}
    \mathbf{a}=\dot{\mathbf{v}}=\frac{\partial \mathbf{v}}{\partial
    t}+\mathbf{\nabla}_{\mathbf{v}}\mathbf{v}.
\end{equation}
%-----------------------------
In components
%-----------------------------
\begin{equation}
    a^a=\frac{\partial v^a}{\partial t}+\frac{\partial v^a}{\partial x^b}v^b+\gamma^a_{bc}v^bv^c.
\end{equation}
%-----------------------------

Let $\varphi:\mathcal{B}\rightarrow \mathcal{S}$ be a $C^1$
configuration of $\mathcal{B}$ in $\mathcal{S}$, where
$\mathcal{B}$ and $\mathcal{S}$ are manifolds. Recall that the
deformation gradient is the tangent map of $\varphi$ and is
denoted by $\mathbf{F}=T\varphi$. Thus, at each point $\mathbf{X}
\in \mathcal{B}$, it is a linear map
%-----------------------------
\begin{equation}
   \mathbf{F}(\mathbf{X}):T_{\mathbf{X}}\mathcal{B}\rightarrow
    T_{\varphi(\mathbf{X})}\mathcal{S}.
\end{equation}
%-----------------------------
If $\{x^a\}$ and $\{X^A\}$ are local coordinate charts on
$\mathcal{S}$ and $\mathcal{B}$, respectively, the components of
$\mathbf{F} $ are
%-----------------------------
\begin{equation}
    F^a{}_{A}(\mathbf{X})=\frac{\partial \varphi^a}{\partial X^A}(\mathbf{X}).
\end{equation}
%-----------------------------
The deformation gradient may be viewed as a two-point tensor
%-----------------------------
\begin{equation}
   \mathbf{F}({\mathbf{X}}):T_{\mathbf{x}}^*\mathcal{S} \times T_{\mathbf{X}}\mathcal{B}\rightarrow \mathbb{R} ; \quad(\mathbf{\alpha},
   \mathbf{V})\mapsto \langle\mathbf{\alpha},T_{\mathbf{X}}\varphi \cdot
   \mathbf{V}\rangle.
\end{equation}
%-----------------------------
Suppose $\mathcal{B}$ and $\mathcal{S}$ are Riemannian manifolds
with inner products $\left\langle \! \left\langle, \right\rangle
\! \right\rangle_{\mathbf{X}}$ and $\left\langle \! \left\langle,
\right\rangle \! \right\rangle_{\mathbf{x}}$ based at
${\mathbf{X}} \in \mathcal{B}$ and ${\mathbf{x}} \in \mathcal{S}$,
respectively. Recall that the transpose of  $\mathbf{F}$ is
defined by
%-----------------------------
\begin{equation}
    \mathbf{F}^{\textsf{T}}:T_{\mathbf{x}}\mathcal{S}  \rightarrow T_{\mathbf{X}}\mathcal{B},~~~
    \left\langle \! \left\langle \mathbf{FV},\mathbf{v}\right\rangle \! \right\rangle_{\mathbf{x}}= \left\langle \!\! \left\langle
    \mathbf{V},\mathbf{F}^{\textsf{T}}\mathbf{v}\right\rangle \!\!
    \right\rangle_{\mathbf{X}},
    \end{equation}
%-----------------------------
for all $\mathbf{V} \in
    T_{\mathbf{X}}\mathcal{B},~\mathbf{v} \in T_{\mathbf{x}} \mathcal{S}$.
In components
%-----------------------------
\begin{equation}
    (F^{\textsf{T}}(\mathbf{X}))^A{}_{a}=g_{ab}(\mathbf{x})F^b{}_{B}(\mathbf{X})G^{AB}(\mathbf{X}),
\end{equation}
%-----------------------------
where $\mathbf{g}$ and $\mathbf{G}$ are metric tensors on
$\mathcal{S}$ and $\mathcal{B}$, respectively. On the other hand,
the {\bfi dual} of $\mathbf{F}$, a metric independent notion, is
defined by
%-----------------------------
\begin{equation}
  \mathbf{F}^*(\mathbf{x}):T_{\mathbf{x}}^*\mathcal{S}\rightarrow T_{\mathbf{X}}^*\mathcal{B}; \quad  \langle\mathbf{F}^*(\mathbf{x})\cdot\alpha,\mathbf{W}\rangle
  =\langle\alpha,\mathbf{F}(\mathbf{X})\mathbf{W}\rangle,
\end{equation}
%-----------------------------
for all $\alpha\in T^*_{\mathbf{x}}\mathcal{S},\mathbf{W}\in
T_{\mathbf{X}}\mathcal{B}$. Considering bases $\mathbf{e}_a$ and
$\mathbf{E}_A$ for $\mathcal{S}$ and $\mathcal{B}$, respectively,
one can define the corresponding dual bases $\mathbf{e}^a$ and
$\mathbf{E}^A$. The matrix representation of $\mathbf{F}^*$ with
respect to the dual bases is the transpose of $F^a{}_{A}$.
$\mathbf{F}$ and $\mathbf{F}^*$ have the following local
representations
%-----------------------------
\begin{equation}
    \mathbf{F}=F^a{}_{A}\frac{\partial}{\partial x^a}\otimes
    dX^A,~~~~~\mathbf{F}^*=F^a{}_{A}dX^A\otimes\frac{\partial}{\partial
    x^a}.
\end{equation}
%-----------------------------
The {\bfi right Cauchy-Green deformation tensor} is defined by
%-----------------------------
\begin{equation}
    \mathbf{C}(X):T_{\mathbf{X}} \mathcal{B}\rightarrow T_{\mathbf{X}} \mathcal{B},~~~~~\mathbf{C}(\mathbf{X})=\mathbf{F}(\mathbf{X})^{\textsf{T}} \mathbf{F}(\mathbf{X}).
\end{equation}
%-----------------------------
In components
%-----------------------------
\begin{equation}
    C^A_{~B}=(F^{\textsf{T}})^A{}_{a}F^a{}_{B}.
\end{equation}
%-----------------------------
It is straightforward to show that
%-----------------------------
\begin{equation}
    \mathbf{C}^\flat=\varphi^*(\mathbf{g})=\mathbf{F}^*\mathbf{g}\mathbf{F},~\textrm{i.e.}~~~C_{AB}=(g_{ab}\circ
    \varphi)F^a{}_{A}F^b{}_{B}.
\end{equation}
%-----------------------------

Let $\varphi_t:\mathcal{B}\rightarrow \mathcal{S}$ be a regular
motion of $\mathcal{B}$ in $\mathcal{S}$ and $\mathcal{P} \subset
\mathcal{B}$ a $p$-dimensional submanifold. The {\bfi Transport
Theorem} says that for any $p$-form $\alpha$ on $\mathcal{S}$
%-----------------------------
\begin{equation}
    \frac{d}{dt}\int_{\varphi_t(\mathcal{P})}\mathbf{\alpha}=
    \int_{\varphi_t(\mathcal{P})}\mathbf{L}_\mathbf{v}\mathbf{\alpha},
\end{equation}
%-----------------------------
where $\mathbf{v}$ is the spatial velocity of the motion. In a
special case when $\mathbf{\alpha}=f dv$ and
$\mathcal{P}=\mathcal{U}$ is an open set, one can write
%-----------------------------
\begin{equation}
    \frac{d}{dt}\int_{\varphi_t(\mathcal{P})}f dv=
    \int_{\varphi_t(\mathcal{P})}\left[\frac{\partial f}{\partial t}+\operatorname{div}
    (f\mathbf{v})\right]dv.
\end{equation}
%-----------------------------

\end{document}